\documentclass[twocolumn,showpacs,preprintnumbers,amsmath,amssymb]{revtex4}
\usepackage{graphicx}% Include figure files
\usepackage{dcolumn}% Align table columns on decimal point
\usepackage{bm}% bold math

\usepackage{color}

\usepackage{amssymb}

\usepackage{epsfig,enumerate}

\newcommand{\sU}{{\sf U}}

\newcommand{\nn}{\nonumber}

\newcommand{\pa}        {\parallel}
\newcommand{\ep}        {\varepsilon}
\newcommand{\la}{{\langle}}
\newcommand{\ra}{{\rangle}}

\newcommand{\ds}{\displaystyle}

\newcommand{\be}{\begin{equation}}
\newcommand{\ee}{\end{equation}}
\newcommand{\ba}{\begin{array}}
\newcommand{\ea}{\end{array}}

\newcommand{\er}[1]{\hbox{(\ref{#1})}}

\newtheorem{theorem}            {Theorem}[section]

\newtheorem{sideremark}         [theorem]{Remark}
\newtheorem{sideeg}           [theorem]{Example}
\newtheorem{sideconj}           [theorem]{Conjecture}
\newtheorem{sideassumption}   [theorem]{Assumption}

\newcommand{\E}                 {{\bf E}}

\newcommand{\bydef}             {\stackrel{\triangle}{=}}

\begin{document}           % End of preamble and beginning of text.

\title{Stability, Gain,  and Robustness  in Quantum Feedback Networks}

\author{C.~D'Helon} \author{M.R.~James}
\affiliation{Department of Engineering, Australian National University,
Canberra, ACT 0200,  Australia.}

\date{\today}
%%%%%%%\maketitle                 % Produces the title.

\begin{abstract}
This paper concerns the problem  of stability for quantum feedback networks. We demonstrate in the context of quantum
optics how  stability of quantum feedback networks  can be guaranteed using only simple gain inequalities for network
components and algebraic relationships determined by the network. Quantum  feedback networks are shown to be stable if
the loop gain is less than one---this is an extension of the famous small gain theorem of classical control theory. We
illustrate the simplicity and power of the small gain approach with applications to  important problems of robust
stability and robust stabilization.

\end{abstract}


\maketitle

%\tableofcontents  % just to help in writing the paper

\section{Introduction}
\label{sec:intro}

Stable operation is a fundamental pre-requisite for the proper functioning of any technological system. Instability
can cause some system variables to grow in magnitude without bound (or at least saturate or oscillate), with
%consequent
detrimental effects on performance and even damage.  Consequently, methods for stability analysis and design have
played an important role in the development of classical technologies. A significant  early example was Watt's steam
engine governor in the 1780's (subsequently analyzed by Maxwell in 1868), \cite{AM96}. Indeed, one of the chief
applications of feedback (but by no means the only application) is to stabilize systems that would otherwise be
unstable. A striking example of this is the X29 plane \cite{GS03}, which has forward-swept wings and requires the use
of a stabilizing feedback control system.

However, feedback per se does not guarantee stability: indeed, feedback can be \lq\lq{degenerative or
regenerative---either stabilizing or destabilizing}\rq\rq, \cite{GZ66a}.
%?? The oscillations in a laser are obtained using feedback.  ??
 In particular, when interconnections of stable
components include components with active elements, instability can occur (such as when a microphone is placed too
close to a loudspeaker). An additional requirement of considerable practical importance is that stable operation be
maintained in the presence of uncertainty (e.g., due to model error and approximation, altered operating conditions,
etc.) and noise---this is a basic robustness requirement.
% The sensitivity and continuity properties of systems and feedback loops are also important considerations, which have
%been explored by Zames, who developed a notion of incremental gain, but we do not discuss them here.

Feedback is increasingly being used in the design of new technologies that include quantum components, e.g.
\cite{VPB83,WM93a,MW98,DJ99,SL00,EDB01,GSM04,HSM05,YK98}. In fact, a wide range of quantum technologies can be
considered as networks of quantum and classical components which include cascade (feedforward)  \cite[chapter 12]{GZ00}
and feedback interconnections. Since these networks may include components that are active, e.g. optical amplifiers or
classical amplifiers, questions of network stability are of considerable importance. Quantum input-output theory
started developing in the 1980's \cite{CG84,CG85}, however general methods for stability analysis and design for
quantum networks still do not appear to be readily available in the literature. The purpose of this paper is to begin
to address this gap; in particular, we show how the {\em small gain theorem} developed in the classical context by
Sandberg, Zames and others in the early 1960's (see, e.g., \cite{VID93,ZDG96,GZ66a}) can be employed with ease and
considerable power for stability analysis and design of quantum networks.

There are many methods for stability analysis available for classical systems, as a glance at any control textbook will
confirm, \cite{OLRJ96,VID93,ZDG96}. For example, it is well known that linear continuous time invariant  classical systems are
asymptotically stable if all the poles of their transfer functions have negative real parts. The Nyquist criteria and
root locus methods are widely used for stability analysis in simple single input single output (SISO) linear
systems. Lyapunov methods are extremely powerful and apply when detailed state space (phase space) models (either
linear or nonlinear) are available. The state-space representation uses a differential equation model of the system
under consideration. A complementary approach, which employs input-output representations,  treats network elements as black boxes,
and describes the relationship between the inputs and outputs of each element. The small gain methods
%\cite{GZ66a,GZ66b}
focus on the input-output properties of systems and form the basis for much of the work that has
been done on robust control system design. The focus on signals entering and leaving network components rather than on
the details of each component makes this technique extremely valuable for analyzing complex feedback networks.

In the small gain feedback stability framework, each component system (of the type shown in Figure
\ref{fig:classical1}) is stable in the sense that bounded input signals produce bounded output signals. This concept of
input-output stability is quantified using the notion of a (time-invariant) {\em gain} $g$. Roughly, if $\pa x \pa$ is
a measure of the \lq\lq{size}\rq\rq \ of a time dependent signal  $x(t)$, we say that the system is {\em bounded input
bounded output (BIBO) stable} if $\pa y \pa \leq g \pa u \pa$, where $u$ and $y$ are respectively the input and output
signals. Gains $g$ less than one correspond to attenuation, while gains greater than one mean amplification. When BIBO
stable components are interconnected in a feedback network, such as in the prototypical example network shown in Figure
\ref{fig:classical2}, it is of importance to know when the network is BIBO stable, considered as a single system, with
respect to inputs $u_0$, $y_0$ and outputs $u_1$, $y_1$, $u_2$, $y_2$, the internal network signals. The {\em small
gain theorem} says that the network will be  stable if the {\em loop gain} is less than one (the loop gain is the
product of all the component gains going around the loop). The loop gain condition is therefore a sufficient but not in
general necessary (due to the absence of phase information) criterion for stability.

In this paper  we consider feedback networks of simple elements taken from quantum optics to illustrate the underlying
principles of the small gain methodology in an important  quantum technology setting. Our aim is to demonstrate how the
small gain  theorem  can guarantee the stability of a complex quantum network using only simple gain inequalities for
network components and algebraic relationships determined by the network. The quantum networks may include classical
components. Loop gain analysis techniques are simple and powerful, and we can expect these and other  methods for
stability analysis and design to have many applications in future quantum technologies. One such application is to {\em
robust stability}, which refers to the ability of a feedback network to remain stable in the presence of uncertainty,
noise, and environmental influences (which may cause decoherence). Here, the environment is considered as a possibly
active network component, parts of which may include unknown model errors, unmodelled dynamics, and noise sources.

The models we use are quantum stochastic differential equations \cite{KRP92,GZ00}, which provide excellent
approximations to quantum optical systems and offer considerable power, with  clear conceptions of input and output
fields. The complete network and noise sources are described by an overall unitary evolution (interaction picture with
respect to the noises). Signals are viewed in \lq\lq{ball and stick}\rq\rq \ terms, \cite{BR04}, and the signal size is
described in mean square terms (average length).

The paper is organized as follows. Section \ref{sec:classical} provides a review of the small gain theorem for
classical systems, both deterministic and stochastic.  As preparation for the small gain analysis of quantum optical
networks, a discussion of the signals and components to be used is given in section \ref{sec:signals-components}, with
particular emphasis on  mean square gains. The main ideas concerning the small gain methodology for quantum networks
are described in section \ref{sec:qsgt}. Then in section \ref{sec:q-robust} we give two examples illustrating important
applications of the small gain theorem to robust stability analysis and design. In particular, we show in the second
example how feedback can be used to increase robustness, so that  the effect of environmental influences is reduced.

\section{Classical Systems}
\label{sec:classical}

This section reviews the meaning of mean square gain for  classical  systems  and summarizes the content of the small
gain theorem for a classical feedback network.

\subsection{Gain for Classical Systems}
\label{sec:c-gain}

\begin{figure}[h]
\begin{center}
\setlength{\unitlength}{1973sp}%
\begingroup\makeatletter\ifx\SetFigFont\undefined%
\gdef\SetFigFont#1#2#3#4#5{%
  \reset@font\fontsize{#1}{#2pt}%
  \fontfamily{#3}\fontseries{#4}\fontshape{#5}%
  \selectfont}%
\fi\endgroup%
\begin{picture}(6624,2124)(3589,-3973)
\thinlines
{\color[rgb]{0,0,0}\put(5401,-3961){\framebox(3000,2100){}}
}%
{\color[rgb]{0,0,0}\put(3601,-2911){\vector( 1, 0){1800}}
}%
{\color[rgb]{0,0,0}\put(8401,-2911){\vector( 1, 0){1800}}
}%
\put(4126,-2686){\makebox(0,0)[lb]{\smash{{\SetFigFont{6}{7.2}{\familydefault}{\mddefault}{\updefault}{\color[rgb]{0,0,0}$u$}%
}}}}
\put(9001,-2686){\makebox(0,0)[lb]{\smash{{\SetFigFont{6}{7.2}{\familydefault}{\mddefault}{\updefault}{\color[rgb]{0,0,0}$y$}%
}}}}
\put(6676,-2911){\makebox(0,0)[lb]{\smash{{\SetFigFont{6}{7.2}{\familydefault}{\mddefault}{\updefault}{\color[rgb]{0,0,0}$\Sigma_c$}%
}}}}
\end{picture}%

\caption{A classical system with input $u$ and  output $y$.}
\label{fig:classical1}
\end{center}
\end{figure}
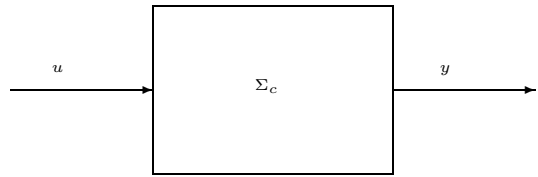

Consider the classical system $\Sigma_c$ shown in Figure \ref{fig:classical1}, with input $u$ and output $y$. The system could be nonlinear. The signals $u$ and $y$ are vector valued functions of time. Here we consider an {\em input-output} description, which does not include internal details; the system is an operator or function mapping input signals to output signals.

The system $\Sigma_c$ is said to have {\em mean square gain} $g > 0$ if $g$ is finite and there exists a constant $\mu
\geq 0$ (called the {\em bias}) such that
\begin{eqnarray}
\int_0^t \vert y(s) \vert^2 ds \leq \mu + g^2 \int_0^t \vert u(s) \vert^2 ds
\label{gain-c-1}
\end{eqnarray}
for all input signals that are mean square finite (square integrable $u \in L^2[0,t]$) on the time interval $[0,t]$, and this should hold  for all $t \geq 0$.
If the system has an internal state variable $x$, then the bias is a function of the initial state $x_0$ in which case we may write $\mu(x_0)$.

The significance of the mean square gain property \er{gain-c-1} is that it captures the important BIBO stability
property of the system, \cite[chapter 6]{VID93}, \cite[chapter 4]{SS99}. In particular, if any mean square finite
signal is applied to the system, then the system responds with a mean square finite output ($u \in L^2$ implies $y \in
L^2$).  If the system has an internal state $x$, then with additional properties like observability or detectability
the stability of $x(t)$ as $t\to\infty$ can be inferred.

\subsection{The Classical Small Gain Theorem}
\label{sec:csgt}

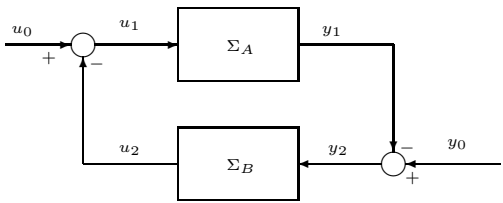
\begin{figure}[h]
\begin{center}
\setlength{\unitlength}{1973sp}%
\begingroup\makeatletter\ifx\SetFigFont\undefined%
\gdef\SetFigFont#1#2#3#4#5{%
  \reset@font\fontsize{#1}{#2pt}%
  \fontfamily{#3}\fontseries{#4}\fontshape{#5}%
  \selectfont}%
\fi\endgroup%
\begin{picture}(6249,2424)(3814,-3073)
\put(4876,-1411){\makebox(0,0)[lb]{\smash{{\SetFigFont{6}{7.2}{\familydefault}{\mddefault}{\updefault}{\color[rgb]{0,0,0}$-$}%
}}}}
{\color[rgb]{0,0,0}\thinlines
\put(4801,-1111){\circle{300}}
}%
{\color[rgb]{0,0,0}\put(6001,-1561){\framebox(1500,900){}}
}%
{\color[rgb]{0,0,0}\put(6001,-3061){\framebox(1500,900){}}
}%
{\color[rgb]{0,0,0}\put(10051,-2611){\vector(-1, 0){1200}}
}%
{\color[rgb]{0,0,0}\put(6001,-2611){\line(-1, 0){1200}}
\put(4801,-2611){\vector( 0, 1){1350}}
}%
{\color[rgb]{0,0,0}\put(4951,-1111){\vector( 1, 0){1050}}
}%
{\color[rgb]{0,0,0}\put(7501,-1111){\line( 1, 0){1200}}
\put(8701,-1111){\vector( 0,-1){1350}}
}%
{\color[rgb]{0,0,0}\put(8551,-2611){\vector(-1, 0){1050}}
}%
{\color[rgb]{0,0,0}\put(3826,-1111){\vector( 1, 0){825}}
}%
\put(6601,-1186){\makebox(0,0)[lb]{\smash{{\SetFigFont{6}{7.2}{\familydefault}{\mddefault}{\updefault}{\color[rgb]{0,0,0}$\Sigma_A$}%
}}}}
\put(6601,-2686){\makebox(0,0)[lb]{\smash{{\SetFigFont{6}{7.2}{\familydefault}{\mddefault}{\updefault}{\color[rgb]{0,0,0}$\Sigma_B$}%
}}}}
\put(5251,-886){\makebox(0,0)[lb]{\smash{{\SetFigFont{6}{7.2}{\familydefault}{\mddefault}{\updefault}{\color[rgb]{0,0,0}$u_1$}%
}}}}
\put(9376,-2386){\makebox(0,0)[lb]{\smash{{\SetFigFont{6}{7.2}{\familydefault}{\mddefault}{\updefault}{\color[rgb]{0,0,0}$y_0$}%
}}}}
\put(8851,-2836){\makebox(0,0)[lb]{\smash{{\SetFigFont{6}{7.2}{\familydefault}{\mddefault}{\updefault}{\color[rgb]{0,0,0}$+$}%
}}}}
\put(8776,-2461){\makebox(0,0)[lb]{\smash{{\SetFigFont{6}{7.2}{\familydefault}{\mddefault}{\updefault}{\color[rgb]{0,0,0}$-$}%
}}}}
\put(3901,-961){\makebox(0,0)[lb]{\smash{{\SetFigFont{6}{7.2}{\familydefault}{\mddefault}{\updefault}{\color[rgb]{0,0,0}$u_0$}%
}}}}
\put(7876,-2461){\makebox(0,0)[lb]{\smash{{\SetFigFont{6}{7.2}{\familydefault}{\mddefault}{\updefault}{\color[rgb]{0,0,0}$y_2$}%
}}}}
\put(5251,-2461){\makebox(0,0)[lb]{\smash{{\SetFigFont{6}{7.2}{\familydefault}{\mddefault}{\updefault}{\color[rgb]{0,0,0}$u_2$}%
}}}}
\put(7801,-961){\makebox(0,0)[lb]{\smash{{\SetFigFont{6}{7.2}{\familydefault}{\mddefault}{\updefault}{\color[rgb]{0,0,0}$y_1$}%
}}}}
\put(4276,-1336){\makebox(0,0)[lb]{\smash{{\SetFigFont{6}{7.2}{\familydefault}{\mddefault}{\updefault}{\color[rgb]{0,0,0}$+$}%
}}}}
{\color[rgb]{0,0,0}\put(8701,-2611){\circle{300}}
}%
\end{picture}%

\caption{A classical feedback network with inputs $u_0$, $y_0$ and internal network signals $u_1$, $u_2$, $y_1$, $y_2$.}
\label{fig:classical2}
\end{center}
\end{figure}

Consider the classical feedback network shown in Figure \ref{fig:classical2}. The network  has inputs $u_0$, $y_0$ and internal network signals $u_1$, $u_2$, $y_1$, $y_2$ (the term {\em internal} here refers to the network, not the internal details of each constituent component system). The classical components $A$ and $B$ are of the type $\Sigma_c$ and satisfy the mean square gain inequality \er{gain-c-1} with gains $g_A$, $g_B$ and biases $\mu_A$, $\mu_B$, respectively.

We are interested in the {\em internal stability} of the network in the BIBO sense, meaning that mean square bounded
input signals should produce mean square bounded internal signals. The {\em small gain theorem}, \cite{GZ66a},
\cite[Theorem 6.6.1-1]{VID93}, \cite[Theorem 4.15]{SS99}, addresses this question, and asserts that the network will be
internally stable  if the {\em loop gain} is strictly smaller than one. That is, if \be g_A g_B < 1 ,
\label{loop-gain-1} \ee then
\begin{eqnarray}
&(1-g_A g_B)\ds \int_0^t \vert z(s) \vert^2 ds
\nn \\
&\leq c_1 + c_2 \ds\int_0^t (\vert u_0(s) \vert^2 + \vert y_0(s) \vert^2 )ds
\label{sgt-c-1}
\end{eqnarray}
for suitable positive constants $c_1$ and $c_2$. Here, $z$ is any of the internal signals $u_1$, $u_2$, $y_1$, $y_2$.
Note that inequality \er{sgt-c-1} provides a meaningful bound on the internal signals only when the loop gain condition
\er{loop-gain-1} holds. Inequality \er{sgt-c-1} is a quantification of the BIBO stability of the feedback network. 
In this section, we have only explicitly considered a network with two elements, but this generalizes in a
straightforward way to any number of elements; similarly the results in this paper also apply more generally. Indeed,  the small-gain theorem applies to multiple input multiple output (MIMO) systems, so that in general the
signals $u,y$ are signal vectors.

\subsection{The Classical Stochastic Case}
\label{sec:classical-stoch}

In the case where $\Sigma_c$ is a stochastic system and $u$ and $y$ are random signals, the mean square gain property
\er{gain-c-1} can be written in terms of expectations
\begin{eqnarray}
\E [\int_0^t \vert y(s) \vert^2 ds]  \leq \E[  \mu  + \lambda t + g^2 \int_0^t \vert u(s) \vert^2 ds ] \label{gain-c-2}
\end{eqnarray}
where $\lambda \geq 0$ is a non-negative constant (related to the variance of the signals). The small gain theorem also
applies to a network of stochastic systems and guarantees internal stability under the same loop gain condition
\er{loop-gain-1}, \cite[section 3]{DJP00}. In this situation, as $t\to\infty$ we must divide the inequality
\er{gain-c-2} by $t$ to obtain a bound on the output signal power in terms of input signal power.

In inequality \er{gain-c-2} we have used $\E$ to denote classical expectation. However, in the remainder of the paper
we use the notation $\la \cdot \ra$ for both classical and quantum expectations.

\section{Quantum Optical Network Signals and Components}
\label{sec:signals-components}

In this paper we are interested in feedback interconnections of quantum systems,  such as the fully quantum feedback loop of Figure \ref{fig:qqfb1}, or the quantum-classical network of Figure \ref{fig:qcfb1}. The purpose of this section is to provide a careful description of the signals and components in these networks. This description will focus on the input-output relations that are needed to facilitate the stability analyses given in  subsequent sections. Standard models will be used to derive these relations. This material may be familiar to some readers.

Electromagnetic fields will be used as the basic carriers of quantum information between quantum network components. We
use the quantum stochastic models as in \cite{GZ00}, \cite{KRP92} to describe the fields, which in some situations
below also serve to model heat baths.  The models are defined on a Hilbert space capturing all components and signals.
Dynamics are described by a unitary operator $\sU(t)$, so that if $X$ is a system operator of a network component, it
evolves in time according to $X(t) = \sU^\dagger(t)X\sU(t)$, which solves a quantum Langevin equation (QLE), see, e.g.,
\cite[equation (11.2.33)]{GZ00}.
%The Hamiltonian of the full network can be obtained using the methods of \cite{YK03a}, though here we use QLEs.
System operators for distinct components will commute ($[X_1(t), X_2(t)]=\sU^\dagger(t)[X_1,X_2]\sU(t)=0$), and satisfy
the non-demolition condition \cite{BB91}, \cite{GZ00}, \cite{LB04} \be [B_{noise}(t), X ] = [B_{noise}^\dagger(t), X
]=0, \ \forall \  t \geq 0, \label{bqnd} \ee where $B_{noise}$ is any purely input noise term in the network or applied
to the network.

Typically, internal quantum signals (e.g. of the form  \er{g-amp-1} below) will be comprised of a zero-mean noise term,
and another term denoted $\beta(t)$, which may contain system operators from network components. The operator function
$\beta(t)$ in general satisfies commutation relations of the form \be [ \beta(t), \beta^\dagger(t)] = c ,
\label{field-displaced-com} \ee for a suitable number $c$.

We use a common   notation for quadratures:
\be
\beta_r = \beta + \beta^\ast, \ \ \beta_i = \frac{\beta-\beta^\ast}{i} .
\label{quad-beta}
\ee
The commutation relation \er{field-displaced-com} implies $[\beta_r, \beta_i]=2ci$, and
\be
\vert \beta(t) \vert^2 \bydef  \beta_r^2(t) + \beta_i^2(t) = 4 \beta^\dagger(t) \beta(t) + 2c ,
\label{quad-beta-1}
\ee
where the (rectangular) modulus notation on the left hand side is defined by the right. In the special case that $\beta$ is a complex number, $\vert \beta \vert^2 = \beta_r^2 + \beta_i^2 = 4 \beta^\ast \beta$, a consequence of the convention \er{quad-beta}.

We often need to take expectations of the various quantities appearing in the networks. Expectations will always be taken with respect to the full state. In general, we do not make explicit the state in the notation. We will also use classical signals, such as electric currents, when classical components are used.

\subsection{Signals and Fields}
\label{sec:components-fields}

\subsubsection{Vacuum Fields}
\label{sec:fields-vacuum}

Calculations (such as expectations) involving  fields in this paper will be all be carried out relative to underlying
vacuum fields (more generally squeezed states with positive temperature could also be considered, though we don't here
for simplicity). These fields will be represented by annihilation operators $b(t)$, which satisfy the canonical
commutation relations (CCR) \be [b(t), b^\dagger(t')] =  \delta(t-t') . \label{ccr-1} \ee We write \be B(t)=\int_0^t
b(s)ds , \ee and use the Ito sense increment $dB(t) = B(t+dt)-B(t)=b(t)dt$. In the vacuum state $\vert 0 \ra$, all Ito
products are zero except $dB(t)dB^\dagger(t)=dt$, \cite[chapter 11]{GZ00}.

The real and imaginary quadratures of the field are defined as
\begin{eqnarray}
Q(t) &=& B(t)+ B^\dagger(t),
\nn \\
P(t) &=& \frac{B(t)- B^\dagger(t)}{i} .
\label{quad-def}
\end{eqnarray}
Both of these quadratures have zero mean and variance $t$ (Ito's rule here reads $dQ(t)dQ(t)=dt$, $dP(t)dP(t)=dt$, $dQ(t)dP(t)=idt$, $dP(t)dQ(t)=-idt$).

Each noise quadrature is self-adjoint (e.g. $Q^\dagger(t) = Q(t)$), and self-commutative (e.g. $[Q(t), Q(s)]=0$ for all
$s,t$). As a consequence, each quadrature is individually stochastically equivalent to a classical Wiener process,
\cite[section 5.2.1]{AH01}. The quadratures $Q(t)$ and $P(t)$, of course, do not commute.

%?? equiv classical

\subsubsection{Coherent and Displaced Fields}
\label{sec:fields-coherent}

A coherent state $\vert \beta \ra$ of a field  is characterized by  a classical complex-valued function of time
$\beta(t)$.  The coherent state is determined by $\vert \beta \ra = W(\beta) \vert 0 \ra_{in}$, where $W(\beta)$ is the
Weyl displacement operator, \cite{KRP92}, \cite{LB04}. To facilitate simple computations involving QLEs, the field in
the coherent state is equivalent to an effective field $b_{\rm{eff}}$  displaced relative to a field operator $b(t)$ in
the vacuum state, and so we write
 \be
 B_{\rm{eff}}(t) = \int_0^t \beta(s)ds + B(t) ,
 \label{g-amp-1}
 \ee
see, e.g., \cite[equation (9.2.47)]{GZ00}.

More generally, we will need to consider fields of the form \er{g-amp-1}, but  with quantum operator $\beta$, as
mentioned above. Note that \er{g-amp-1} is a basic  \lq\lq{signal or message  plus noise\rq\rq \  model, or \lq\lq{ball
and stick}\rq\rq \  model, \cite{BR04} (semimartingale  in mathematical terms).

The field quadratures are defined as in \er{quad-def}, and satisfy
\begin{eqnarray}
dQ_{\rm{eff}}(t) = \beta_r(t)dt + dQ(t) ,
\nn \\
dP_{\rm{eff}}(t) = \beta_i(t)dt + dP(t) , \label{quad-in}
\end{eqnarray}
where $\beta_r$ and $\beta_i$ are given by \er{quad-beta}.

The root mean square size of the field $b_{\rm{eff}}$ on a time interval $[0,t]$ is defined to be
\be \pa\beta\pa_t = \sqrt{\int_0^t \la \vert
\beta(s) \vert^2 \ra ds} , \label{q-ms}
\ee
which for an optical field is a measure of the intensity. The quantum
expectation takes into account the fluctuations due to quantum noise.

\subsubsection{Classical Signals}

Classical signals, such as currents and voltages, occur in electro-optical networks. These signals are of the
\lq\lq{ball and stick}\rq\rq \ form \be dy(t)= \beta_y(t)dt + dw_y(t) , \label{sig-c} \ee where $\beta_y(t)$ is a
classical stochastic process and $w_y(t)$ is a Wiener process. This signal has root mean square size
\be \pa \beta_y
\pa_t = \sqrt{\int_0^t \la \vert \beta_y(s) \vert^2 \ra ds} . \label{c-ms}
\ee
on a time interval $[0,t]$. We remind the reader that we use $\la
\cdot \ra$ to denote both quantum and classical expectations.

%?? mean square

\subsection{Some Quantum Network Components and their Gains}
\label{sec:components}

In this section we review  some components  that are commonly used in quantum optics, and pay particular attention to their gain properties. These devices will be used in the networks considered later in the paper.

\subsubsection{Beamsplitters}
\label{sec:components-bs}

A beamsplitter is shown in Figure \ref{fig:bs1}. The input-output relations are \cite[section 5.1]{BR04}
\begin{eqnarray}
b_{out,1} &=&  \ep b_{in,1} - \delta b_{in,2}
\nn \\
b_{out,2} &=&  \delta b_{in,1} + \ep b_{in,2} ,
\label{bs1-1}
\end{eqnarray}
where
\be
\ep^2 + \delta^2 = 1 .
\ee
%In the sequel we take for simplicity
%\be
%\ep = \delta = \frac{1}{\sqrt{2}} .
%\ee
The parameters $\ep$ and $\delta$ describe the levels of transmission and attenuation for the field channels.
Note that the output fields satisfy the CCRs \er{ccr-1}.

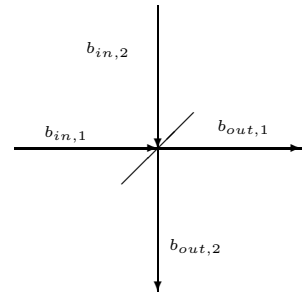
\begin{figure}[h]
\begin{center}
\setlength{\unitlength}{1973sp}%
\begingroup\makeatletter\ifx\SetFigFont\undefined%
\gdef\SetFigFont#1#2#3#4#5{%
  \reset@font\fontsize{#1}{#2pt}%
  \fontfamily{#3}\fontseries{#4}\fontshape{#5}%
  \selectfont}%
\fi\endgroup%
\begin{picture}(3624,3624)(3439,-5323)
\put(5401,-4786){\makebox(0,0)[lb]{\smash{{\SetFigFont{6}{7.2}{\familydefault}{\mddefault}{\updefault}{\color[rgb]{0,0,0}$b_{out,2}$}%
}}}}
\thinlines
{\color[rgb]{0,0,0}\put(3451,-3511){\vector( 1, 0){1800}}
}%
{\color[rgb]{0,0,0}\put(5251,-3511){\vector( 1, 0){1800}}
}%
{\color[rgb]{0,0,0}\put(5251,-1711){\vector( 0,-1){1800}}
}%
{\color[rgb]{0,0,0}\put(5251,-3511){\vector( 0,-1){1800}}
}%
\put(3826,-3361){\makebox(0,0)[lb]{\smash{{\SetFigFont{6}{7.2}{\familydefault}{\mddefault}{\updefault}{\color[rgb]{0,0,0}$b_{in,1}$}%
}}}}
\put(6001,-3286){\makebox(0,0)[lb]{\smash{{\SetFigFont{6}{7.2}{\familydefault}{\mddefault}{\updefault}{\color[rgb]{0,0,0}$b_{out,1}$}%
}}}}
\put(4351,-2311){\makebox(0,0)[lb]{\smash{{\SetFigFont{6}{7.2}{\familydefault}{\mddefault}{\updefault}{\color[rgb]{0,0,0}$b_{in,2}$}%
}}}}
{\color[rgb]{0,0,0}\put(4801,-3961){\line( 1, 1){900}}
}%
\end{picture}%

\caption{A beamsplitter with inputs $b_{in,1}$, $b_{in,2}$ and outputs  $b_{out,1}$, $b_{out,2}$.}
\label{fig:bs1}
\end{center}
\end{figure}

%?? bs gain

\subsubsection{Cavities}
\label{sec:components-cav}

We  consider resonant optical cavities, with a simple first order dynamical model \cite[section 5.3]{BR04},
\cite[chapter 12]{GZ00}, \cite[section III]{YK03a}. The cavity is weakly coupled to a resonant external field, with
parameter $\gamma$, about a nominal frequency. The QLE for the cavity annihilation operator $a(t)$ in the interaction
picture is \be da(t) = -\frac{\gamma}{2} a(t) dt -\sqrt{\gamma}  dB_{in}(t) . \label{cavity1-1} \ee

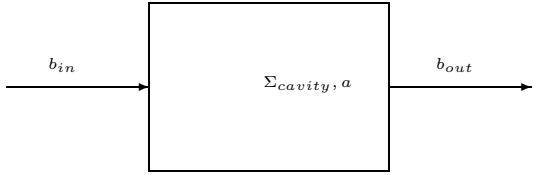
\begin{figure}[h]
\begin{center}
\setlength{\unitlength}{1973sp}%
\begingroup\makeatletter\ifx\SetFigFont\undefined%
\gdef\SetFigFont#1#2#3#4#5{%
  \reset@font\fontsize{#1}{#2pt}%
  \fontfamily{#3}\fontseries{#4}\fontshape{#5}%
  \selectfont}%
\fi\endgroup%
\begin{picture}(6624,2124)(3589,-3973)
\thinlines
{\color[rgb]{0,0,0}\put(5401,-3961){\framebox(3000,2100){}}
}%
{\color[rgb]{0,0,0}\put(3601,-2911){\vector( 1, 0){1800}}
}%
{\color[rgb]{0,0,0}\put(8401,-2911){\vector( 1, 0){1800}}
}%
\put(4126,-2686){\makebox(0,0)[lb]{\smash{{\SetFigFont{6}{7.2}{\familydefault}{\mddefault}{\updefault}{\color[rgb]{0,0,0}$b_{in}$}%
}}}}
\put(9001,-2686){\makebox(0,0)[lb]{\smash{{\SetFigFont{6}{7.2}{\familydefault}{\mddefault}{\updefault}{\color[rgb]{0,0,0}$b_{out}$}%
}}}}
\put(6826,-2911){\makebox(0,0)[lb]{\smash{{\SetFigFont{6}{7.2}{\familydefault}{\mddefault}{\updefault}{\color[rgb]{0,0,0}$\Sigma_{cavity},a$}%
}}}}
\end{picture}%

\caption{A network representation of an optical cavity with annihilation operator $a$, showing input $b_{in}$ and
output $b_{out}$ fields.} \label{fig:cavity1}
\end{center}
\end{figure}

%If $\sU(t)$ denotes the unitary evolution of the total system, then $a(t)=\sU^\dagger(t) a(0) \sU(t)$ and %$B_{out}(t) = \sU^\dagger(t) B_{in}(t) \sU(t)$.

%?? The QSDE for $\sU(t)$ is ???? if needed ???

If the input field $b_{in}$ is taken to be a displaced field, of the form \er{g-amp-1}, with operator-valued $\beta(t)$
\be B_{in}(t) = \int_0^t \beta(s)ds + B(t) , \label{amp-1-in} \ee i.e., \be dB_{in}(t) = \beta(t)dt + dB(t) , \nn \ee
then the output field is given by
\begin{eqnarray}
dB_{out}(t) &=&  \sqrt{\gamma} a(t) dt + dB_{in}(t)
\nn \\
&=& \beta_{out}(t) dt + dB(t)
\label{cavity1-2}
\end{eqnarray}
where the output operator $\beta_{out}$ is given by
\be
\beta_{out}(t) = \sqrt{\gamma} a(t) + \beta(t) .
\label{cavity1-2a}
\ee

Consider the cavity quadratures
\begin{eqnarray*}
q = a+ a^\dagger, \ \
p =  \frac{a-a^\dagger}{i} ,
\end{eqnarray*}
and the output field quadratures
\begin{eqnarray}
dQ_{out}(t) &=&  \beta_{out,r}(t) + dQ(t),
\nn \\
dP_{out}(t) &=&  \beta_{out,i}(t) + dP(t),
\label{cavity1-2aa}
\end{eqnarray}
where
\begin{eqnarray}
\beta_{out,r}(t) &=&  \sqrt{\gamma} q(t) + \beta_r(t),
\nn \\
\beta_{out,i}(t) &=&  \sqrt{\gamma} p(t) + \beta_i(t) .
\label{cavity1-2ab}
\end{eqnarray}

Then
\begin{eqnarray}
dq(t) &=& (-\frac{\gamma}{2} q(t) - \sqrt{\gamma} \beta_r(t))dt - \sqrt{\gamma} dQ(t) ,
\label{g-cav-13}
\end{eqnarray}
and similarly,
\begin{eqnarray}
dp(t) = ( -\frac{\gamma}{2} p(t) - \sqrt{\gamma}\beta_i(t)) dt - \sqrt{\gamma} dP(t) .
\label{g-cav-15}
\end{eqnarray}

Consider now  the mean squares of cavity quadratures. By completion of squares in the above we have
\begin{eqnarray}
dq^2(t) &=&  (dq(t) )q(t)+ q(t) dq(t) + dq(t) dq(t)
\nn  \\
&=& (- \beta_{out,r}^2(t) +  \beta_r^2(t) + \gamma ) dt
\nn \\
&& \ - 2\sqrt{\gamma} q(t) dQ(t) .
\label{g-cav-20}
\end{eqnarray}
Combining this with the analogous expression for $dp^2(t)$, and taking the integral of the expectations gives
\begin{eqnarray}
\la q^2(t)+ p^2(t)  \ra + \int_0^t \la \vert \beta_{out}(s) \vert^2 \ra ds \hspace{1.0cm}
\label{g-cav-21} \\
= \la q^2(0)+ p^2(0) \ra +   \int_0^t \la \vert \beta(s) \vert^2 \ra ds + \lambda t ,
\nn
\end{eqnarray}
where $ \lambda = 2\gamma $ (recall the modulus notation \er{quad-beta-1}). This is equivalent to a statement of energy
conservation for the cavity, and implies \be \pa \beta_{out} \pa_t ^2 \leq \mu_0 + \lambda t + \pa \beta \pa_t ^2 ,
\label{gain-q-2} \ee where $\mu_0 = \la q^2(0)+ p^2(0) \ra$.
%\begin{eqnarray}
%& \ds \int_0^t \la \vert \beta_{out}(s) \vert^2 \ra \ra ds \hspace{1.0cm}
%\label{gain-q-2} \\
%&\leq  \la q^2(0)+ p^2(0) \ra +  \ds\int_0^t \la \vert \beta(s) \vert^2  \ra ds + \lambda t .
%\nn
%\end{eqnarray}
This inequality is of the form of the classical gain inequality \er{gain-c-2}, with gain $g=1$.
%A detailed cavity model can be used to determine a stricter gain value.
%We close this subsection with a stronger bound on the mean square values of the cavity quadratures:
%\begin{eqnarray}
%& \la q^2(t) + p^2(t) \ra \leq e^{-\frac{\gamma}{2}t} \la q^2(0) +p^2(0) \ra
%\nn
%\\
%& + C \ds\int_0^t e^{-\frac{\gamma}{2}(t-s)} \la  \vert  \beta(s) \vert^2 \ra ds ,
%\label{g-cav-23}
%\end{eqnarray}
%for some positive constant $C$. The derivation of this bound uses the variation of constants formula for
%linear differential equations, and to handle cross terms the  following bound was used:
%for any  numbers $a$ and $b$, we have
%\be \vert a + b \vert^2 \leq (1+\eta^{-1}) a^2 + (1+\eta)b^2 , \label{qc-sgt-2} \ee for $\forall \eta > 0$.

\subsubsection{Amplifiers and Attenuators}
\label{sec:components-amps}

An optical amplifier or attenuator is  shown in Figure \ref{fig:amp1}. We consider first the amplifier, and consider
its gain. We use the inverted temperature model described in \cite[chapter 7]{GZ00} for a linear amplifier. The model
consists of a gain medium in an optical cavity, described by an inverted heat bath $b_{aux}$, which provides the gain.
The input field $b_{in}$ is also coupled to the cavity mode and an amplified output field $b_{out}$ is produced. Of
necessity, this process introduces noise, as documented in \cite[chapter 7]{GZ00} and \cite{CC82}. In general, this
includes additive noise due to fluctuations in the inverted heat bath
%(i.e., due to spontaneous emission)
which is independent of the input signal, as well as multiplicative noise which amplifies the fluctuations in the input
signal.

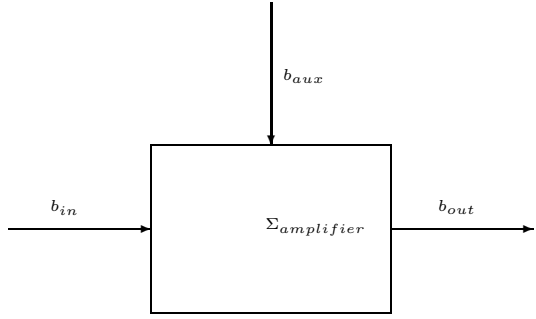
\begin{figure}[h]
\begin{center}
\setlength{\unitlength}{1973sp}%
\begingroup\makeatletter\ifx\SetFigFont\undefined%
\gdef\SetFigFont#1#2#3#4#5{%
  \reset@font\fontsize{#1}{#2pt}%
  \fontfamily{#3}\fontseries{#4}\fontshape{#5}%
  \selectfont}%
\fi\endgroup%
\begin{picture}(6624,3924)(3589,-3973)
\put(6826,-2911){\makebox(0,0)[lb]{\smash{{\SetFigFont{6}{7.2}{\familydefault}{\mddefault}{\updefault}{\color[rgb]{0,0,0}$\Sigma_{amplifier}$}%
}}}}
\thinlines
{\color[rgb]{0,0,0}\put(3601,-2911){\vector( 1, 0){1800}}
}%
{\color[rgb]{0,0,0}\put(8401,-2911){\vector( 1, 0){1800}}
}%
{\color[rgb]{0,0,0}\put(6901,-61){\vector( 0,-1){1800}}
}%
\put(4126,-2686){\makebox(0,0)[lb]{\smash{{\SetFigFont{6}{7.2}{\familydefault}{\mddefault}{\updefault}{\color[rgb]{0,0,0}$b_{in}$}%
}}}}
\put(9001,-2686){\makebox(0,0)[lb]{\smash{{\SetFigFont{6}{7.2}{\familydefault}{\mddefault}{\updefault}{\color[rgb]{0,0,0}$b_{out}$}%
}}}}
\put(7051,-1036){\makebox(0,0)[lb]{\smash{{\SetFigFont{6}{7.2}{\familydefault}{\mddefault}{\updefault}{\color[rgb]{0,0,0}$b_{aux}$}%
}}}}
{\color[rgb]{0,0,0}\put(5401,-3961){\framebox(3000,2100){}}
}%
\end{picture}%

\caption{An optical amplifier or attenuator with gain $g$, showing input $b_{in}$ and output $b_{out}$ fields, and the
auxiliary input $b_{aux}$.} \label{fig:amp1}
\end{center}
\end{figure}

For an input field $b_{in}$ of the form \er{amp-1-in}, the amplifier model is described by \be da(t) = \frac{\gamma}{2}
a(t) dt -\sqrt{\gamma}  dB_{aux}(t) -\frac{\kappa}{2} a(t) dt -\sqrt{\kappa}  dB_{in}(t) \label{amp1-1} \ee and \be
dB_{out}(t) = \beta_{out}(t) dt + dB(t) , \label{amp1-1e} \ee where \be \beta_{out}(t) = \sqrt{\kappa} a(t) + \beta(t)
. \label{amp1-1b} \ee The energy required for the gain is provided by the inverted heat bath field $b_{aux}$, which has
non-zero Ito product \be dB^\dagger_{aux}(t) dB_{aux}(t) = dt . \label{amp1-1c} \ee We assume $\kappa > \gamma > 0$ for
stable amplifier operation.

Calculations analogous to those in subsection \ref{sec:components-cav} show that
\begin{eqnarray}
d \la q^2(t) + p^2(t) \ra &=& (\la -(\kappa-\gamma) ( q^2(t)+p^2(t))
\label{amp2-1} \\
&&  - \sqrt{\kappa}( \beta_r(t) q(t) + q(t)\beta_r(t)
\nn \\
&& + \beta_i(t) p(t) + p(t) \beta_i(t) \ra + \lambda )dt ,
\nn
\end{eqnarray}
where $\lambda = 2(\kappa + \gamma)$. Define
$$
\tilde{\beta}_{out}(t) = \sqrt{\kappa - \gamma} \, a(t) + \sqrt{\frac{\kappa}{\kappa - \gamma}} \beta(t) ,
$$
which is related to the output operator $\beta_{out}$ by
\be
\beta_{out}(t) = \sqrt{\frac{\kappa}{\kappa - \gamma}} \tilde \beta_{out}(t) - \frac{\gamma}{\kappa-\gamma} \beta(t) .
\label{amp2-3}
\ee
Then by completion of squares \er{amp2-1} implies
\begin{eqnarray}
& \ds \int_0^t \la \vert \tilde\beta_{out}(s) \vert^2 \ra \ra ds \hspace{1.0cm}
\label{amp2-2} \\
&\leq  \la q^2(0)+ p^2(0) \ra +  \ds{ \frac{\kappa}{\kappa - \gamma}  \int_0^t \la \vert \beta(s) \vert^2  \ra ds} + \lambda t ,
\nn
\end{eqnarray}
and using the relation \er{amp2-3} and some calculations we arrive at the mean square gain inequality c.f.,
\er{gain-c-2} \be \pa \beta_{out} \pa_t ^2 \leq \mu + \lambda t + g^2 \pa \beta \pa_t ^2 , \label{amp2-2a} \ee
%\begin{eqnarray}
%& \ds \int_0^t \la \vert \beta_{out}(s) \vert^2 \ra \ra ds \hspace{1.0cm}
%\label{amp2-2} \\
%&\leq  \mu_0 + g^2  \ds{  \int_0^t \la \vert \beta(s) \vert^2  \ra ds} + \lambda_0 t ,
%\nn
%\end{eqnarray}
where the gain $g$ is given by
\be
g =  \frac{\kappa+\gamma}{\kappa - \gamma} > 1 ,
 \label{amp2-4}
 \ee
and $\mu$ and $\lambda$ are suitable constants, \footnote{The gain $g$ is given in \cite[eq. (7.2.13)]{GZ00} derived
using a steady-state approximation. It also equals the $H^\infty$ norm \cite[chapter ??]{ZDG96} of the amplifier
frequency response $G(i\omega)=(\gamma+\kappa - 2i\omega)/(\gamma-\kappa-2i\omega)$, \cite[eq. (7.2.17)]{GZ00}.}. The
inequality \er{amp2-2a} is an upper bound on the size of the output signal, with the additional noise amplified by
$\nu=\sqrt{g^2-1}$. We note that a lower bound on the output signal has been derived previously e.g., \cite{CC82},
which highlights the quantum limit on noise in linear amplifiers.

Attenuators can be analyzed similarly. The auxiliary field $b_{aux}$ is a standard (non-inverted) heat bath, to
facilitate loss. A mean square gain inequality of the form \er{amp2-2a} also holds, but with gain \be g =  \ds \vert
\frac{\gamma - \kappa}{\gamma + \kappa} \vert < 1,
 \label{amp2-5}
 \ee
and noise amplification $\nu=\sqrt{1-g^2}$. When, as here, the gain is smaller than one, attenuation occurs.

We mention also that idealized static models for amplifiers and attenuators are described in \cite[chapter 7]{GZ00},
and take the form \be b_{out}(t) = g b_{in}(t) - \nu b_{aux}(t) , \label{amp1-1a} \ee where \be g^2 + \sigma \nu^2 =1 ,
\label{amp1-2a} \ee and \be \sigma = \left\{ \ba{ll} +1 & \  \text{if} \ 0 < g < 1, \ \text{(attenuator)}
\\
-1 & \ \text{if} \ g > 1, \     \text{(amplifier)} \ea \right. \label{amp1-3a} \ee In the case that $g=1$, this model
reduces to that for a resonant optical cavity.

\subsubsection{Quadrature Measurement}
\label{sec:components-hd}

A schematic representation of the measurement of the real quadrature $Q_{in}(t)$ of the field $B_{in}(t)$ given by
\er{g-amp-1} is shown in Figure \ref{fig:quad1}.

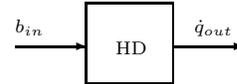
\begin{figure}[h]
\begin{center}
\setlength{\unitlength}{2368sp}%
\begingroup\makeatletter\ifx\SetFigFont\undefined%
\gdef\SetFigFont#1#2#3#4#5{%
  \reset@font\fontsize{#1}{#2pt}%
  \fontfamily{#3}\fontseries{#4}\fontshape{#5}%
  \selectfont}%
\fi\endgroup%
\begin{picture}(2424,849)(4339,-1198)
\thinlines
{\color[rgb]{0,0,0}\put(5101,-1186){\framebox(900,825){}}
}%
{\color[rgb]{0,0,0}\put(6001,-811){\vector( 1, 0){750}}
}%
{\color[rgb]{0,0,0}\put(4351,-811){\vector( 1, 0){750}}
}%
\put(5401,-886){\makebox(0,0)[lb]{\smash{{\SetFigFont{7}{8.4}{\familydefault}{\mddefault}{\updefault}{\color[rgb]{0,0,0}HD}%
}}}}
\put(4351,-661){\makebox(0,0)[lb]{\smash{{\SetFigFont{7}{8.4}{\familydefault}{\mddefault}{\updefault}{\color[rgb]{0,0,0}$b_{in}$}%
}}}}
\put(6226,-661){\makebox(0,0)[lb]{\smash{{\SetFigFont{7}{8.4}{\familydefault}{\mddefault}{\updefault}{\color[rgb]{0,0,0}$\dot q_{out}$}%
}}}}
\end{picture}%

\caption{Homodyne detection of the real quadrature $q_{in}$ of an input field $b_{in}$ produces a current $\dot q_{out}$.}
\label{fig:quad1}
\end{center}
\end{figure}

The detection scheme (homodyne) produces a photocurrent $\dot q_{out}$, which can be described by the Ito equation \be
dq_{out}(t) =  \tilde  \beta_r(t)  dt +  d\tilde w(t), \label{hd-out-1} \ee where $\tilde w(t)$ is a standard Wiener
process and $\tilde \beta_r(t)$ is related to the real quadrature $\beta_r(t)$ in the input field \er{g-amp-1}. The
basic inequality for root mean square values before (for the entire field $\beta$) and after (just for the measured
quadrature $\tilde \beta_r$) measurement is \be \pa \tilde\beta_r \pa_t \leq \pa \beta \pa_t . \label{hd-id-2} \ee
%\begin{eqnarray}
%\int_0^t \la \tilde \beta_r^2(s)  \ra  ds  \leq \int_0^t  \la \vert \beta(s) \vert^2 \ra  ds. \label{hd-id-2}
%\end{eqnarray}
Hence the gain of the quadrature measurement is not more than one.

We remark that the photocurrent can be represented in two ways. We consider $B_{in}(t)$ to be an output of some quantum
component. The quadrature $Q_{in}(t)$ is a self-commutative quantum  stochastic process, statistically equivalent to
the classical quantity $q_{out}(t)$. The driving noise quadrature $Q(t)$  is also self-commutative, and statistically
equivalent to a standard classical Wiener processes.  So $\tilde w(t)$ and $\tilde \beta_r(t)$ can be interpreted as
the classical statistical equivalents of $dQ$ and $\beta_r$. This gives the first of the two representations. The
second representation depends on quantum filtering and stochastic master equation considerations, and views $\tilde
w(t)$ as the innovation process, also a standard Wiener process but distinct from the one in the first interpretation.
In this case, the process $\tilde \beta_r(t)$ is a quantity $\check \beta_r(t)$ obtained from a conditional}
 stochastic master equation.\footnote{$\check \beta_r(t)$ is the {\em conditional expectation} of $\beta_r(t)$, \cite[section
3.4]{BGM04}.}

\subsubsection{Modulators}

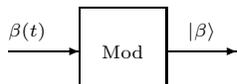
\begin{figure}[h]
\begin{center}
\setlength{\unitlength}{2368sp}%
\begingroup\makeatletter\ifx\SetFigFont\undefined%
\gdef\SetFigFont#1#2#3#4#5{%
  \reset@font\fontsize{#1}{#2pt}%
  \fontfamily{#3}\fontseries{#4}\fontshape{#5}%
  \selectfont}%
\fi\endgroup%
\begin{picture}(2424,849)(4339,-1198)
\thinlines
{\color[rgb]{0,0,0}\put(5101,-1186){\framebox(900,825){}}
}%
{\color[rgb]{0,0,0}\put(6001,-811){\vector( 1, 0){750}}
}%
{\color[rgb]{0,0,0}\put(4351,-811){\vector( 1, 0){750}}
}%
\put(4351,-661){\makebox(0,0)[lb]{\smash{{\SetFigFont{7}{8.4}{\familydefault}{\mddefault}{\updefault}{\color[rgb]{0,0,0}$\beta(t)$}%
}}}}
\put(6226,-661){\makebox(0,0)[lb]{\smash{{\SetFigFont{7}{8.4}{\familydefault}{\mddefault}{\updefault}{\color[rgb]{0,0,0}$\vert \beta \ra$}%
}}}}
\put(5326,-886){\makebox(0,0)[lb]{\smash{{\SetFigFont{7}{8.4}{\familydefault}{\mddefault}{\updefault}{\color[rgb]{0,0,0}Mod}%
}}}}
\end{picture}%

\caption{An electro-optical modulator produces a field $b_{out}$ in a coherent state $\vert \beta \ra$ (with intensity
$|\beta|^2$) from a classical signal $\beta(t)$.} \label{fig:mod1}
\end{center}
\end{figure}

The final component we use in this paper is the modulator, shown in Figure \ref{fig:mod1}, \cite[section 5.4.3]{BR04}.

We shall assume that the modulator has gain no more than one: \be \pa \beta \pa_{t,(out)} \leq \pa \beta \pa_{t,(in)} .
\label{mod-1} \ee
%\begin{eqnarray}
%\ds \int_0^t \la \vert  \beta(s)  \vert^2 \ra ds \leq   \ds \int_0^t \la \vert  \beta_{(c)}(s) \vert^2 \ra ds . \
%\label{mod-1}
%\end{eqnarray}

\section{The Small Gain Methodology for Quantum Optical Networks}
\label{sec:qsgt}

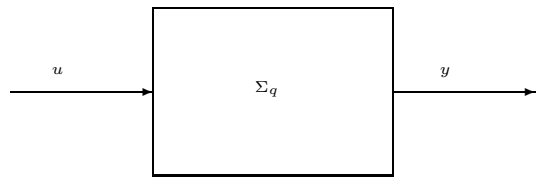
\begin{figure}[h]
\begin{center}
\setlength{\unitlength}{1973sp}%
\begingroup\makeatletter\ifx\SetFigFont\undefined%
\gdef\SetFigFont#1#2#3#4#5{%
  \reset@font\fontsize{#1}{#2pt}%
  \fontfamily{#3}\fontseries{#4}\fontshape{#5}%
  \selectfont}%
\fi\endgroup%
\begin{picture}(6624,2124)(3589,-3973)
\thinlines
{\color[rgb]{0,0,0}\put(5401,-3961){\framebox(3000,2100){}}
}%
{\color[rgb]{0,0,0}\put(3601,-2911){\vector( 1, 0){1800}}
}%
{\color[rgb]{0,0,0}\put(8401,-2911){\vector( 1, 0){1800}}
}%
\put(4126,-2686){\makebox(0,0)[lb]{\smash{{\SetFigFont{6}{7.2}{\familydefault}{\mddefault}{\updefault}{\color[rgb]{0,0,0}$u$}%
}}}}
\put(9001,-2686){\makebox(0,0)[lb]{\smash{{\SetFigFont{6}{7.2}{\familydefault}{\mddefault}{\updefault}{\color[rgb]{0,0,0}$y$}%
}}}}
\put(6676,-2911){\makebox(0,0)[lb]{\smash{{\SetFigFont{6}{7.2}{\familydefault}{\mddefault}{\updefault}{\color[rgb]{0,0,0}$\Sigma_q$}%
}}}}
\end{picture}%

\caption{A quantum system with input $u$ and  output $y$.}
\label{fig:quantum1}
\end{center}
\end{figure}

In this section we describe the small gain methodology for stability analysis to quantum optical networks.  The
procedure is to first determine the gain of each of the components, and the algebraic relationships among the signals.
Each component could be of the form shown in Figure \ref{fig:quantum1}, with inputs and outputs being displaced signals
of the form
\begin{eqnarray}
dU(t) &=& \beta_u(t)dt + dB_u(t)
\label{general-u} \\
dY(t) &=& \beta_y(t)dt + dB_y(t)
\label{general-y}
\end{eqnarray}

What is important is a mean square gain inequality of the form
 \be \pa \beta_y \pa_t ^2 \leq \mu + \lambda t + g^2 \pa
\beta_u \pa_t ^2 ,
\label{gain-q-1}
\ee
%\begin{eqnarray}
%& \ds\int_0^t \la \vert \beta_y(s) \vert^2 \ra ds \nn \\
%& \leq \mu + g^2 \ds\int_0^t \la \vert  \beta_u(s) \vert^2 \ra ds + \lambda t ,
%\label{gain-q-1}
%\end{eqnarray}
which emphasizes input-output properties of the components.  The dynamics of the component are unspecified, and need
not be linear - the only stipulation is that the dynamics are those of a valid quantum system. In general, the
non-negative number $\lambda$ depends on the noise variances. Note that components may be active elements, or may
dissipate energy, and while  connections to external heat baths would be involved, these are not necessarily shown
explicitly.

For a closed network the loop gain is the product of the gains going around a loop. We will demonstrate in this section
that if the loop gain is less than one then all internal network signals are mean square bounded in terms of the mean
square inputs. The feedback network is then called {\em internally stable}.

If one is also interested in the behavior of internal component variables (system operators), then additional information is needed.  This information is often available from explicit physical models.

\subsection{A Quantum Feedback Network}
\label{sec:qo-coherent}

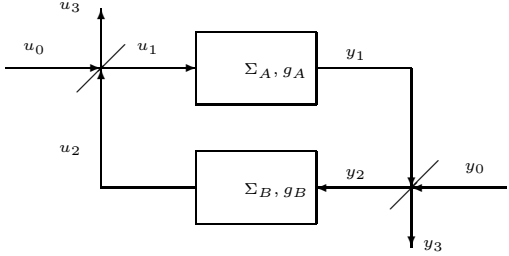
\begin{figure}[h]
\begin{center}
\setlength{\unitlength}{1973sp}%
\begingroup\makeatletter\ifx\SetFigFont\undefined%
\gdef\SetFigFont#1#2#3#4#5{%
  \reset@font\fontsize{#1}{#2pt}%
  \fontfamily{#3}\fontseries{#4}\fontshape{#5}%
  \selectfont}%
\fi\endgroup%
\begin{picture}(6324,3214)(3589,-3419)
\put(7876,-2461){\makebox(0,0)[lb]{\smash{{\SetFigFont{6}{7.2}{\familydefault}{\mddefault}{\updefault}{\color[rgb]{0,0,0}$y_2$}%
}}}}
\thinlines
{\color[rgb]{0,0,0}\put(6001,-3061){\framebox(1500,900){}}
}%
{\color[rgb]{0,0,0}\put(5101,-811){\line(-1,-1){600}}
}%
{\color[rgb]{0,0,0}\put(9038,-2273){\line(-1,-1){600}}
}%
{\color[rgb]{0,0,0}\put(4801,-2611){\line( 1, 0){1200}}
}%
{\color[rgb]{0,0,0}\put(7501,-1111){\line( 1, 0){1200}}
}%
{\color[rgb]{0,0,0}\put(4801,-1111){\vector( 1, 0){1200}}
}%
{\color[rgb]{0,0,0}\put(4801,-2611){\vector( 0, 1){1500}}
}%
{\color[rgb]{0,0,0}\put(8701,-1111){\vector( 0,-1){1500}}
}%
{\color[rgb]{0,0,0}\put(8701,-2611){\vector(-1, 0){1200}}
}%
{\color[rgb]{0,0,0}\put(8701,-2611){\vector( 0,-1){750}}
}%
{\color[rgb]{0,0,0}\put(4801,-1111){\vector( 0, 1){750}}
}%
{\color[rgb]{0,0,0}\put(3601,-1111){\vector( 1, 0){1200}}
}%
{\color[rgb]{0,0,0}\put(9901,-2611){\vector(-1, 0){1200}}
}%
\put(6601,-1186){\makebox(0,0)[lb]{\smash{{\SetFigFont{6}{7.2}{\familydefault}{\mddefault}{\updefault}{\color[rgb]{0,0,0}$\Sigma_A, g_A$}%
}}}}
\put(6601,-2686){\makebox(0,0)[lb]{\smash{{\SetFigFont{6}{7.2}{\familydefault}{\mddefault}{\updefault}{\color[rgb]{0,0,0}$\Sigma_B, g_B$}%
}}}}
\put(3826,-886){\makebox(0,0)[lb]{\smash{{\SetFigFont{6}{7.2}{\familydefault}{\mddefault}{\updefault}{\color[rgb]{0,0,0}$u_0$}%
}}}}
\put(5251,-886){\makebox(0,0)[lb]{\smash{{\SetFigFont{6}{7.2}{\familydefault}{\mddefault}{\updefault}{\color[rgb]{0,0,0}$u_1$}%
}}}}
\put(4276,-2161){\makebox(0,0)[lb]{\smash{{\SetFigFont{6}{7.2}{\familydefault}{\mddefault}{\updefault}{\color[rgb]{0,0,0}$u_2$}%
}}}}
\put(4276,-361){\makebox(0,0)[lb]{\smash{{\SetFigFont{6}{7.2}{\familydefault}{\mddefault}{\updefault}{\color[rgb]{0,0,0}$u_3$}%
}}}}
\put(9376,-2386){\makebox(0,0)[lb]{\smash{{\SetFigFont{6}{7.2}{\familydefault}{\mddefault}{\updefault}{\color[rgb]{0,0,0}$y_0$}%
}}}}
\put(8851,-3361){\makebox(0,0)[lb]{\smash{{\SetFigFont{6}{7.2}{\familydefault}{\mddefault}{\updefault}{\color[rgb]{0,0,0}$y_3$}%
}}}}
\put(7876,-961){\makebox(0,0)[lb]{\smash{{\SetFigFont{6}{7.2}{\familydefault}{\mddefault}{\updefault}{\color[rgb]{0,0,0}$y_1$}%
}}}}
{\color[rgb]{0,0,0}\put(6001,-1561){\framebox(1500,900){}}
}%
\end{picture}%

\caption{A fully quantum feedback loop, consisting of two quantum components with gains $g_A$ and $g_B$ linked by two
beamsplitters.} \label{fig:qqfb1}
\end{center}
\end{figure}

We consider now the stability of the fully quantum feedback loop shown in Figure \ref{fig:qqfb1}, consisting of two
components with gains $g_A$, $g_B$, respectively, linked by two beamsplitters with attenuation parameters $\delta_A$,
$\delta_B$, and transmissivity parameters $\ep_A$, $\ep_B$ respectively. The small gain technique will be used to show
that the network is {\em internally stable} when the loop gain is strictly less than one. The meaning of this statement
will become clear in what follows.

\subsubsection{Network Inputs and Internal Signals}
\label{sec:qo-coherent-in}

The input fields $u_0$, $y_0$  are displaced fields   given by
\begin{eqnarray}
U_0(t) &=&  \int_0^t \beta_{u_0}(s)ds + B_{u_0}(t) ,
\label{qc-1} \\
Y_0(t) &=&  \int_0^t \beta_{y_0}(s)ds + B_{y_0}(t) ,
\label{qq-1}
\end{eqnarray}
where $\beta_{u_0}$, $\beta_{y_0}$ are operator-valued and $B_{u_0}$, $B_{y_0}$ are vacuum  fields.

The internal network signals, $u_1,u_2,y_1,y_2$, will be displaced signals of the form \er{g-amp-1}, and we use
notation analogous to \er{qc-1}, \er{qq-1}.

\subsubsection{Small Gain Theorem}
\label{sec:qq-sgt}

We now prove a quantum version of the small gain theorem for the network of Figure \ref{fig:qqfb1}, assuming that the
following loop gain condition holds: \be \delta_A \delta_B g_A g_B < 1 . \label{qq-n-loop-gain-1} \ee The loop gain
here is the product $\delta_A g_A \delta_B g_B$, which takes into account the attenuation due to the beamsplitters. We
shall show that \be \pa \beta \pa_t \leq C\left(1 + \sqrt{t} + \pa \beta_{u_0} \pa_t + \pa \beta_{y_0} \pa_t \right),
\label{qq-sgt-n-1} \ee
%\begin{eqnarray}
%&(1-\delta_A \delta_B g_A g_B)\ds \int_0^t \la  \vert \beta(s) \vert^2  \ra ds \nn \\
%& \leq C(1 + t + \nn
%& +  \ds\int_0^t ( \vert \beta_{u_0}(s) \vert^2  + \vert \beta_{y_0}(s) \vert^2  )ds ) \label{qq-sgt-n-1} \\
%\end{eqnarray}
for some positive constant $C$, where $\beta$ corresponds to any of the internal network signals $\beta_{u_1}$,$
\beta_{y_2}$, $\beta_{y_1}$, $\beta_{u_2}$.

%??? explain BIBO

\subsubsection{Network Analysis}
\label{sec:qo-coherent-eqns}

The beamsplitter equations \er{bs1-1} imply
\begin{eqnarray}
u_1 &=&  \ep_A u_0 - \delta_A u_2
\nn \\
u_3 &=&  \delta_Au_0 + \ep_A u_2 ,
\label{qq-bs-1}
\end{eqnarray}
and
\begin{eqnarray}
y_2 &=&  \ep_B y_0 - \delta_B y_1
\nn \\
y_3 &=&  \delta_B y_0 + \ep_B y_1 .
\label{qq-bs-2}
\end{eqnarray}

We go around the loop clockwise. The signal $u_1$ exiting the top-left beamsplitter gives
\begin{eqnarray}
U_1(t) &=& \ep_A U_0(t) - \delta_A U_2(t)
\label{qq-3}  \\
&=&  \int_0^t  \beta_{u_1}(s) ds + B_{u_1}(t) \nn
\end{eqnarray}
where
\begin{eqnarray}
\beta_{u_1}(t)&=& \ep_A \beta_{u_0}(t) - \delta_A \beta_{u_2}(t),
\label{qq-4} \\
dB_{u_1}(t)&=& \ep_A dB_{u_0}(t) - \delta_A dB_{u_2}(t) .
\label{qq-5}
\end{eqnarray}
Note that $\beta_{u_1}$ is a quantum signal since $\beta_{u_0}$ and $\beta_{u_2}$ are  quantum signals. The output of component $\Sigma_A$
is of the form
\begin{eqnarray}
Y_1(t)
&=& \int_0^t \beta_{y_1}(s)  ds + B_{y_1}(t) ,
\label{qq-6}
\end{eqnarray}
and is related to the input by the gain inequality \be \pa \beta_{y_1} \pa_t ^2 \leq \mu_A + \lambda_A t + g_A^2 \pa
\beta_{u_1} \pa_t ^2 , \label{qq-9} \ee
%\begin{eqnarray}
%\int_0^t \la \vert \beta_{y_1}(s) \vert^2 \ra ds \leq  \mu_A \nn \\ + g_A^2 \int_0^t \la \vert \beta_{u_1}(s) \vert^2
%\ra ds +\lambda_A t . \label{qq-9}
%\end{eqnarray}

Similarly, considering the bottom-right beamsplitter and component $\Sigma_B$ we have \be \pa \beta_{u_2} \pa_t ^2 \leq
\mu_B + \lambda_B t + g_B^2 \pa \beta_{y_2} \pa_t^2 , \label{qq-10} \ee
%\begin{eqnarray}
%\int_0^t \la \vert \beta_{u_2}(s)\vert^2 \ra ds \leq  \mu_B \nn \\ + g_B^2 \int_0^t \la \vert \beta_{y_2 }(s) \vert^2
%\ra ds + \lambda_B t , \label{qq-10}
%\end{eqnarray}
where
\begin{eqnarray}
\beta_{y_2}(t) &=& \ep_B \beta_{y_0} - \delta_B \beta_{y_1}(t) .
\label{qq-12}
\end{eqnarray}

\subsubsection{Small Gain Calculations}

Continuing with the derivation of the small gain bounds, we first
%observe that by \er{qq-n-loop-gain-1} there exists a
%number $\eta > 0$ such that \be (1+\eta)^2 \delta_A  g_A \delta_B g_B < 1 . \label{qc-sgt-1} \ee
consider the field $u_1$. From  \er{qq-4}, and using the triangle inequality
%bound for any real numbers $a$ and $b$,}
%\be \vert a + b \vert^2 \leq (1+\eta^{-1}) a^2 + (1+\eta)b^2 , \forall \eta > 0 \label{qc-sgt-2} \ee
\be
 \pa c_1\beta_1 + c_2\beta_2 \pa_t \leq |c_1| \pa \beta_1 \pa_t + |c_2| \pa \beta_2 \pa_t , \label{qc-ineq-01}
 \ee
  we have \be \pa \beta_{u_1} \pa_t \leq \epsilon_A \pa \beta_{u_0} \pa_t + \delta_A \pa
\beta_{u_2} \pa_t . \label{loop-u1-01} \ee Going around the loop anticlockwise, we consider the field $u_2$. From
\er{qq-10}, and using the inequality \be |c|^2 + g^2 \pa \beta \pa_t^2 \leq \left( |c| + g \pa \beta \pa_t \right)^2 ,
\label{qc-ineq-02} \ee to obtain the root mean square, we have \be \pa \beta_{u_2} \pa_t \leq c_B(t) + g_B \pa
\beta_{y_2} \pa_t , \label{loop-u2-01} \ee where $c_B(t)=\sqrt{\mu_B + \lambda_B t}$. Substituting \er{loop-u2-01} into
\er{loop-u1-01}, \be \pa \beta_{u_1} \pa_t \leq \epsilon_A \pa \beta_{u_0} \pa_t + \delta_A c_B(t) + \delta_A g_B \pa
\beta_{y_2} \pa_t . \label{loop-u1-02} \ee
%\begin{eqnarray}
%& \int_0^t   \la \vert \beta_{u_1}(s) \vert^2   \ra ds
%\nn \\
% & \leq
%\ep_A^2 (1+\eta^{-1}) \ds \int_0^t \la \vert \beta_{u_0} (s) \vert^2 \ra ds
%\nn \\
%& + \delta_A^2 (1+\eta) \ds \int_0^t \la \vert \beta_{u_2} (s) \vert^2 \ra ds
%\nn \\
%& \leq
%\ep_A^2 (1+\eta^{-1})  \ds \int_0^t \la \vert  \beta_{u_0} (s)\vert^2 \ra ds
%\nn \\
%&+ \delta_A^2 (1+\eta) \left( \mu_B + g_B^2   \ds \int_0^t \la \vert \beta_{y_2} (s) \vert^2 \ra ds  +\lambda_B t
%\right)
%\label{qq-sgt-3}
%\end{eqnarray}

Next, we consider the field $y_2$. Using the relations \er{qq-12} and \er{qc-ineq-01} we have \be \pa \beta_{y_2} \pa_t
\leq \epsilon_B \pa \beta_{y_0} \pa_t + \delta_B \pa \beta_{y_1} \pa_t . \label{loop-y2-01} \ee Substituting
\er{loop-y2-01} into \er{loop-u1-02} gives
\begin{eqnarray}
\pa \beta_{u_1} \pa_t & \leq & \delta_A c_B(t) + \epsilon_A \pa \beta_{u_0} \pa_t +
\delta_A \epsilon_B g_B \pa \beta_{y_0} \pa_t \nn \\
& & + \delta_A \delta_B g_B \pa \beta_{y_1} \pa_t . \label{loop-u1-03}
\end{eqnarray}
%\begin{eqnarray}
%& \ds\int_0^t   \la \vert \beta_{y_2}(s) \vert^2   \ra ds
%\nn \\
% & \leq  \ep_B^2 (1+\eta^{-1})  \ds\int_0^t   \la \vert \beta_{y_0}(s)\vert^2   \ra ds
% \nn \\
%& + \delta^2_B(1+\eta) \left(\mu_A + g_A^2 \ds\int_0^t   \la \vert \beta_{u_1}(s)\vert^2   \ra ds + \lambda_A t \right)
%. \label{qq-sgt-4}
%\end{eqnarray}
Finally we consider the field $y_1$. From \er{qq-9} and \er{qc-ineq-02} we have \be \pa \beta_{y_1} \pa_t \leq c_A(t) +
g_A \pa \beta_{u_1} \pa_t , \label{loop-y1-01} \ee where $c_A(t)=\sqrt{\mu_A + \lambda_A t}$. Substituting
\er{loop-y1-01} into \er{loop-u1-03} and rearranging, yields the desired bound \er{qq-sgt-n-1} for field $u_1$,
\begin{eqnarray}
(1-\delta_A \delta_B g_A g_B)\pa \beta_{u_1} \pa_t & \leq & \nn
\delta_A(c_B(t) + \delta_B g_B c_A(t)) \nn \\
& & + \epsilon_A\pa \beta_{u_0} \pa_t \nn \\
& & + \delta_A\epsilon_B g_B \pa \beta_{y_0} \pa_t
\end{eqnarray}
The other internal fields in Figure \ref{fig:qqfb1} can be bounded in a similar manner.

%Combining \er{qq-sgt-3} and \er{qq-sgt-4} we find that
%\begin{eqnarray}
%&(1-\delta_A^2\delta_B^2(1+\eta)^2g_A^2g_B^2)  \ds\int_0^t   \la \vert \beta_{u_1}(s) \vert^2   \ra ds
%\nn \\
%&\leq \ep_A^2 (1+\eta^{-1})  \ds \int_0^t \la \vert \beta_{u_0} (s) \vert^2 \ra ds
%\label{qq-sgt-5} \\
%&+\delta_A^2(1+\eta) \{ \mu_B + g_B^2 ( \ep_B^2 (1+\eta^{-1})  \ds\int_0^t   \la \vert \beta_{y_0}(s) \vert^2   \ra ds
%\nn \\ & +\delta_B^2(1+\eta) \mu_A + (\lambda_B + (1+\eta)\lambda_A)t ) \} \nn.
%\end{eqnarray}

\subsection{A Quantum-Classical Feedback Network}
\label{sec:qc}

\begin{figure}[h]
\begin{center}
\setlength{\unitlength}{1973sp}%
\begingroup\makeatletter\ifx\SetFigFont\undefined%
\gdef\SetFigFont#1#2#3#4#5{%
  \reset@font\fontsize{#1}{#2pt}%
  \fontfamily{#3}\fontseries{#4}\fontshape{#5}%
  \selectfont}%
\fi\endgroup%
\begin{picture}(6474,2868)(3589,-3073)
\put(6301,-2986){\makebox(0,0)[lb]{\smash{{\SetFigFont{6}{7.2}{\familydefault}{\mddefault}{\updefault}{\color[rgb]{0,0,0}classical}%
}}}}
\thinlines
{\color[rgb]{0,0,0}\put(6001,-1561){\framebox(1500,900){}}
}%
{\color[rgb]{0,0,0}\put(6001,-3061){\framebox(1500,900){}}
}%
{\color[rgb]{0,0,0}\put(5101,-811){\line(-1,-1){600}}
}%
{\color[rgb]{0,0,0}\put(4801,-1111){\vector( 1, 0){1200}}
}%
{\color[rgb]{0,0,0}\put(4801,-1111){\vector( 0, 1){750}}
}%
{\color[rgb]{0,0,0}\put(3601,-1111){\vector( 1, 0){1200}}
}%
{\color[rgb]{0,0,0}\put(8551,-2611){\vector(-1, 0){1050}}
}%
{\color[rgb]{0,0,0}\put(10051,-2611){\vector(-1, 0){1200}}
}%
{\color[rgb]{0,0,0}\put(8401,-1411){\framebox(600,600){}}
}%
{\color[rgb]{0,0,0}\put(7501,-1111){\vector( 1, 0){900}}
}%
{\color[rgb]{0,0,0}\put(8701,-1411){\vector( 0,-1){1050}}
}%
{\color[rgb]{0,0,0}\put(4801,-1561){\vector( 0, 1){450}}
}%
{\color[rgb]{0,0,0}\put(4501,-2161){\framebox(600,600){}}
}%
{\color[rgb]{0,0,0}\put(6001,-2611){\line(-1, 0){1200}}
}%
{\color[rgb]{0,0,0}\put(4801,-2611){\vector( 0, 1){450}}
}%
\put(6601,-1186){\makebox(0,0)[lb]{\smash{{\SetFigFont{6}{7.2}{\familydefault}{\mddefault}{\updefault}{\color[rgb]{0,0,0}$\Sigma_A,g_A$}%
}}}}
\put(6601,-2686){\makebox(0,0)[lb]{\smash{{\SetFigFont{6}{7.2}{\familydefault}{\mddefault}{\updefault}{\color[rgb]{0,0,0}$\Sigma_B,g_B$}%
}}}}
\put(5251,-886){\makebox(0,0)[lb]{\smash{{\SetFigFont{6}{7.2}{\familydefault}{\mddefault}{\updefault}{\color[rgb]{0,0,0}$u_1$}%
}}}}
\put(4276,-361){\makebox(0,0)[lb]{\smash{{\SetFigFont{6}{7.2}{\familydefault}{\mddefault}{\updefault}{\color[rgb]{0,0,0}$u_3$}%
}}}}
\put(9376,-2386){\makebox(0,0)[lb]{\smash{{\SetFigFont{6}{7.2}{\familydefault}{\mddefault}{\updefault}{\color[rgb]{0,0,0}$y_0$}%
}}}}
\put(7876,-2386){\makebox(0,0)[lb]{\smash{{\SetFigFont{6}{7.2}{\familydefault}{\mddefault}{\updefault}{\color[rgb]{0,0,0}$y_2$}%
}}}}
\put(8851,-1861){\makebox(0,0)[lb]{\smash{{\SetFigFont{6}{7.2}{\familydefault}{\mddefault}{\updefault}{\color[rgb]{0,0,0}$\tilde y_1$}%
}}}}
\put(8851,-2836){\makebox(0,0)[lb]{\smash{{\SetFigFont{6}{7.2}{\familydefault}{\mddefault}{\updefault}{\color[rgb]{0,0,0}$+$}%
}}}}
\put(8776,-2461){\makebox(0,0)[lb]{\smash{{\SetFigFont{6}{7.2}{\familydefault}{\mddefault}{\updefault}{\color[rgb]{0,0,0}$-$}%
}}}}
\put(3901,-961){\makebox(0,0)[lb]{\smash{{\SetFigFont{6}{7.2}{\familydefault}{\mddefault}{\updefault}{\color[rgb]{0,0,0}$u_0$}%
}}}}
\put(5251,-2461){\makebox(0,0)[lb]{\smash{{\SetFigFont{6}{7.2}{\familydefault}{\mddefault}{\updefault}{\color[rgb]{0,0,0}$\beta_{u_2}$}%
}}}}
\put(4876,-1411){\makebox(0,0)[lb]{\smash{{\SetFigFont{6}{7.2}{\familydefault}{\mddefault}{\updefault}{\color[rgb]{0,0,0}$u_2$}%
}}}}
\put(4576,-1936){\makebox(0,0)[lb]{\smash{{\SetFigFont{6}{7.2}{\familydefault}{\mddefault}{\updefault}{\color[rgb]{0,0,0}Mod}%
}}}}
\put(8476,-1186){\makebox(0,0)[lb]{\smash{{\SetFigFont{6}{7.2}{\familydefault}{\mddefault}{\updefault}{\color[rgb]{0,0,0}HD}%
}}}}
\put(7726,-886){\makebox(0,0)[lb]{\smash{{\SetFigFont{6}{7.2}{\familydefault}{\mddefault}{\updefault}{\color[rgb]{0,0,0}$y_1$}%
}}}}
\put(6301,-1486){\makebox(0,0)[lb]{\smash{{\SetFigFont{6}{7.2}{\familydefault}{\mddefault}{\updefault}{\color[rgb]{0,0,0}quantum}%
}}}}
{\color[rgb]{0,0,0}\put(8701,-2611){\circle{300}}
}%
\end{picture}%

\caption{A quantum-classical feedback loop, consisting of a quantum component $\Sigma_A$ with gain $g_A$ and a
classical (e.g., electronic) device $\Sigma_B$ with gain $g_B$ linked  using opto-electronic connections. The quantum
to classical transition (measurement) is via homodyne detection (HD), while the modulator (Mod) turns the classical
signal $\beta_{u_2}$ into a coherent state $\vert \beta_{u_2} \ra$, shown as the field $u_2$.} \label{fig:qcfb1}
\end{center}
\end{figure}
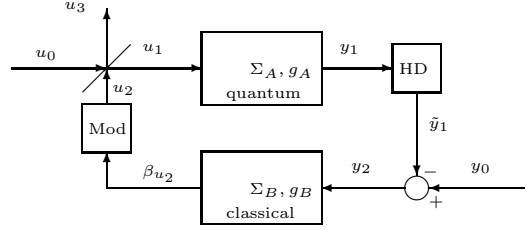

In this section we derive a small gain theorem for the quantum-classical feedback network shown in Figure \ref{fig:qcfb1}. Stability will be assessed relative to a quantum  input field $u_0$ and a classical noise source $y_0$.

\subsubsection{Network Inputs}

We begin by considering the two inputs. The input field $u_0$ is  a displaced field  given by \er{qc-1},
where $B_{u_0}$ is a vacuum  field. The classical input signal $\dot y_0$ is given by
\be
dy_0(t) = \beta_{y_0}(t) dt + dw_{y_0}(t) ,
\label{qc-1-a}
\ee
where $\beta_{y_0}$ is a classical real-valued signal, and $w_{y_0}$ is a standard Wiener process.

\subsubsection{Network Analysis}
\label{sec:qc-net}

Next, we present explicitly the equations and inequalities for the network, working around the loop clockwise,
beginning with the beamsplitter. The output field of the modulator is given by \be U_2(t)=  \int_0^t \beta_{u_2}(s)ds +
B_{u_2}(t) , \label{qc-2} \ee where $\beta_{u_2}(t)$ is  the amplitude of the coherent state $\vert \beta_{u_2} \ra$
created by the classical output of $\Sigma_B$, and $B_{u_2}$ is a vacuum field. The signal $u_1$ from the beamsplitter
is
\begin{eqnarray}
U_1(t) &=& \ep U_0(t) - \delta U_2(t)
\label{qc-3}  \\
&=&  \int_0^t  \beta_{u_1}(s) ds + B_{u_1}(t) \nn
\end{eqnarray}
where
\begin{eqnarray}
\beta_{u_1}(t)&=& \ep \beta_{u_0}(t) - \delta \beta_{u_2}(t),
\label{qc-4} \\
dB_{u_1}(t)&=& \ep dB_{u_0}(t) - \delta dB_{u_2}(t)
\label{qc-5}
\end{eqnarray}
Note that $\beta_{u_1}$ is in general operator-valued. The output of the quantum component $\Sigma_A$ is of the form
\er{qq-6} with gain inequality \er{qq-9}.

Next, from \er{hd-out-1} the output of the quadrature measurement (HD) is the classical signal
\begin{eqnarray}
d\tilde{y}_1(t) = \tilde{\beta}_{y_1,r}(t) dt + dw_1(t) , \label{qc-10}
\end{eqnarray}
where $w_1$ is a standard Wiener process ($dw_1(t)dw_1(t)=dt$). From \er{hd-id-2} the corresponding root mean square
inequality is \be \pa \tilde\beta_{y_1,r} \pa_t \leq \pa \beta_{y_1} \pa_t . \label{qc-11} \ee
%\be
%\int_0^t \la \check \beta_{y_1,r}^2 (s) \ra ds  \leq \int_0^t \la \vert  \beta_{y_1}(s) \vert^2 \ra ds . \label{qc-11}
%\ee

The classical system $\Sigma_B$ has input
\begin{eqnarray}
dy_2(t) &=& - d\tilde y_1(t) + dy_0(t)
\nn \\
&=& \beta_{y_2}(t) dt + dB_{y_2}(t) , \label{qc-14}
\end{eqnarray}
where
\begin{eqnarray}
\beta_{y_2}(t) &=& \beta_{y_0}(t) - \tilde{\beta}_{y_1,r}(t) ,
\label{qc-15}  \\
dB_{y_2}(t) &=& dw_{y_0}(t) - dw_1(t) , \label{qc-16}
\end{eqnarray}
and is assumed to have mean square gain $g_B$: \be \pa \beta_{u_2} \pa_t^2 \leq \mu_B + \lambda_B t + g_B^2 \pa
\beta_{y_2} \pa_t^2 , \label{qc-12} \ee
%\begin{eqnarray}
%&  \ds\int_0^t  \la  \vert \beta_{u_2}(s)\vert^2 \ra  ds
%\nn \\
%& \leq \mu_B + g_B^2  \ds\int_0^t  \la \vert  \beta_{y_2}(s) \vert^2   \ra  ds ] + \lambda_B t . \label{qc-12}
%\end{eqnarray}

The final part of the loop is the modulator (Mod), for which we have \be \pa \beta_{u_2} \pa_{t,(out)} \leq \pa
\beta_{u_2} \pa_{t,(in)} . \label{qc-13} \ee
%\begin{eqnarray}
% \ds \int_0^t \la \vert \beta_{u_2}(s)\vert^2 \ra ds
% \leq   \ds \int_0^t \la \vert  \beta_{u_2,(c)}(s)\vert^2 \ra ds , \
% \label{qc-13}
% \end{eqnarray}
using \er{mod-1}.

\subsubsection{Small Gain Theorem}
\label{sec:qc-sgt}

We will show that the loop gain condition \be \delta g_A g_B < 1 \label{loop-gain-3} \ee implies that if the inputs
$u_0$ and $y_0$ are mean square bounded, then any signal $\beta$ in the feedback loop of Figure \ref{fig:qcfb1} will be
mean square bounded, in the sense that there exists a positive constant $C > 0$ such that \be \pa \beta \pa_t \leq
C\left(1 + \sqrt{t} + \pa \beta_{u_0} \pa_t + \pa \beta_{y_0} \pa_t \right), \label{qq-sgt-0} \ee
%\begin{eqnarray}
% \int_0^t \la \vert \beta(s) \vert^2 \ra ds  \leq
%C (1 + t +
%\nn \\
%\ \ + \int_0^t \la \vert \beta_{u_0}(s)\vert^2 + (\vert \beta_{y_0}(s)\vert^2 \ra ds ) , \label{qc-sgt-0}
%\end{eqnarray}
%where $\beta=\beta_{u_1}, \beta_{y_2}, \beta_{y_1}, \beta_{u_2}$.

%Before using the network relations derived in section \ref{sec:qc-net}, we note that by \er{loop-gain-3} there exists $\eta > 0$ such that
%\be
%(1+\eta)^2 \delta g_A g_B < 1 .
%\label{qc-sgt-1}
%\ee

We first consider the field $u_1$. Using the triangle inequality \er{qc-ineq-01}  and
applying the modulator inequality \er{qc-13}, we have \be \pa \beta_{u_1} \pa_t \leq \epsilon \pa \beta_{u_0} \pa_t +
\delta \pa \beta_{u_2} \pa_t . \label{qc-sgt-3} \ee
%Consider the field $u_1$. From \er{qc-4}, \er{qc-12},  \er{qc-13}
%and using the inequality \er{qc-sgt-2} we have
%\begin{eqnarray}
%\int_0^t \la \vert \beta_{u_1}(s) \vert^2 \ra ds & \leq & \ep^2 (1+\eta^{-1})  \int_0^t \la \vert \beta_{u_0}(s)
%\vert^2 \ra ds
%\nn \\
%&& + \delta^2 (1+\eta) \int_0^t \la \vert \beta_{u_2}(s) \vert^2 \ra ds
%\nn \\
%& \leq &
%\ep^2 (1+\eta^{-1}) \int_0^t \la \vert \beta_{u_0}(s) \vert^2 \ra ds
%\nn \\
%+ \delta^2 (1+\eta)& g_B^2 & \left(  \int_0^t  \la \vert \beta_{y_2}(s) \vert^2 \ra ds + \lambda_B t \right)
%\label{qc-sgt-3}
%\end{eqnarray}

Going around the loop anticlockwise, we consider the classical signal $\beta_{u_2}$. From \er{qc-12}, and using the
inequality \er{qc-ineq-02} to obtain the root mean square, we have \be \pa \beta_{u_2} \pa_t \leq c_B(t) + g_B \pa
\beta_{y_2} \pa_t , \label{loop-u2-c1} \ee where $c_B(t)=\sqrt{\mu_B + \lambda_B t}$. Substituting \er{loop-u2-c1} into
\er{qc-sgt-3}, \be \pa \beta_{u_1} \pa_t \leq \epsilon \pa \beta_{u_0} \pa_t + \delta c_B(t) + \delta g_B \pa
\beta_{y_2} \pa_t . \label{loop-u1-c2} \ee

Next, we consider the classical signal $y_2$. Using the inequality \er{qc-ineq-01} to bound \er{qc-15} we have \be \pa
\beta_{y_2} \pa_t \leq \pa \beta_{y_0} \pa_t + \pa \tilde\beta_{y_1,r} \pa_t . \label{loop-y2-c1} \ee Substituting
\er{loop-y2-c1} into \er{loop-u1-c2}, and applying the measurement inequality \er{qc-11} gives
\begin{eqnarray}
\pa \beta_{u_1} \pa_t & \leq & \delta c_B(t) + \epsilon \pa \beta_{u_0} \pa_t + \delta g_B \pa \beta_{y_0} \pa_t \nn \\
& & + \delta g_B \pa \beta_{y_1} \pa_t . \label{loop-u1-c3}
\end{eqnarray}

Finally we consider the field $y_1$, which is the output of the quantum component $\Sigma_A$, with mean square gain
inequality \er{qq-9}. From \er{qq-9} and the inequality \er{qc-ineq-02} we have \be \pa \beta_{y_1} \pa_t \leq c_A(t) +
g_A \pa \beta_{u_1} \pa_t , \label{loop-y1-c1} \ee where $c_A(t)=\sqrt{\mu_A + \lambda_A t}$. Substituting
\er{loop-y1-c1} into \er{loop-u1-c3} and rearranging, yields the required bound \er{qq-sgt-0} for field $u_1$,
\begin{eqnarray}
(1-\delta g_A g_B)\pa \beta_{u_1} \pa_t & \leq & \nn
\delta(c_B(t) + g_B c_A(t)) \nn \\
& & + \epsilon\pa \beta_{u_0} \pa_t \nn \\
& & + \delta g_B\pa \beta_{y_0} \pa_t .
\end{eqnarray}
The other internal fields in Figure \ref{fig:qcfb1} can be bounded in a similar manner.

%Next, using the relation \er{qc-sgt-2}, \er{qc-15}, \er{qc-11} and \er{qq-9} we have
%\begin{eqnarray}
% \int_0^t  \la \vert \beta_{y_2}(s) \vert^2  \ra ds
% & \leq & (1+\eta^{-1}) \int_0^t \la \vert \beta_{y_0}(s)\vert^2 \ra ds \nn \\
%+ (1+\eta) & & \left( \mu_A + g_A^2 \int_0^t \la  \vert \beta_{u_1}(s)\vert^2 \ra ds + \lambda_A t \right)
%\label{qc-sgt-4} .
%\end{eqnarray}
%Combining \er{qc-sgt-3} and \er{qc-sgt-4} we find that
%\begin{eqnarray}
%(1-\delta^2(1+\eta)^2g_A^2g_B^2)\int_0^t \la \vert \beta_{u_1}(s) \vert^2 \ra ds
%\nn \\
%\leq \ep^2 (1+\eta^{-1})  \int_0^t \la \vert \beta_{u_0}(s) \vert^2 \ra ds \nn \\
%+\delta^2(1+\eta)g_B^2 ( (1+\eta^{-1})\int_0^t \la \vert \beta_{y_0}(s) \vert^2 \ra ds
%\nn \\
%+(1+\eta) \mu_A + \nn \\ (\lambda_B + (1+\eta)\lambda_A)t ) \label{qc-sgt-5} .
%\end{eqnarray}
%This implies \er{qc-sgt-0} for $\beta=\beta_{u_1}$. The remaining cases are handled in a similar way.

\section{Applications to Robust Stability Analysis and Design}
\label{sec:q-robust}

In this section we illustrate two of the important applications of the small gain methodology, namely robust stability
analysis and design. The basic issue addressed is {\em robust stability}. The first example (subsection
\ref{sec:q-robust1}) analyzes the condition on the maximum allowable external disturbance for which a nominal feedback
network remains stable. In the second example (subsection \ref{sec:osc}) we design a controller which improves the
robustness of an open oscillator e.g., an atom trapped in an optical cavity.
%by increasing the maximum allowable disturbance due to the environment.}

\subsection{A Simple Quantum Feedback Loop}
\label{sec:q-robust1}

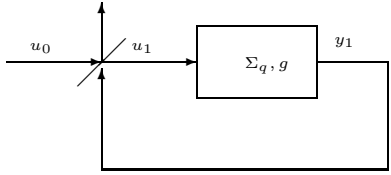
\begin{figure}[h]
\begin{center}
\setlength{\unitlength}{1973sp}%
\begingroup\makeatletter\ifx\SetFigFont\undefined%
\gdef\SetFigFont#1#2#3#4#5{%
  \reset@font\fontsize{#1}{#2pt}%
  \fontfamily{#3}\fontseries{#4}\fontshape{#5}%
  \selectfont}%
\fi\endgroup%
\begin{picture}(4824,2124)(3589,-2473)
\put(5176,-961){\makebox(0,0)[lb]{\smash{{\SetFigFont{6}{7.2}{\familydefault}{\mddefault}{\updefault}{\color[rgb]{0,0,0}$u_1$}%
}}}}
\thinlines
{\color[rgb]{0,0,0}\put(5101,-811){\line(-1,-1){600}}
}%
{\color[rgb]{0,0,0}\put(4801,-1111){\vector( 1, 0){1200}}
}%
{\color[rgb]{0,0,0}\put(4801,-1111){\vector( 0, 1){750}}
}%
{\color[rgb]{0,0,0}\put(3601,-1111){\vector( 1, 0){1200}}
}%
{\color[rgb]{0,0,0}\put(7501,-1111){\line( 1, 0){900}}
\put(8401,-1111){\line( 0,-1){1350}}
\put(8401,-2461){\line(-1, 0){3600}}
\put(4801,-2461){\vector( 0, 1){1275}}
}%
\put(6601,-1186){\makebox(0,0)[lb]{\smash{{\SetFigFont{6}{7.2}{\familydefault}{\mddefault}{\updefault}{\color[rgb]{0,0,0}$\Sigma_q,g$}%
}}}}
\put(3901,-961){\makebox(0,0)[lb]{\smash{{\SetFigFont{6}{7.2}{\familydefault}{\mddefault}{\updefault}{\color[rgb]{0,0,0}$u_0$}%
}}}}
\put(7726,-886){\makebox(0,0)[lb]{\smash{{\SetFigFont{6}{7.2}{\familydefault}{\mddefault}{\updefault}{\color[rgb]{0,0,0}$y_1$}%
}}}}
{\color[rgb]{0,0,0}\put(6001,-1561){\framebox(1500,900){}}
}%
\end{picture}%

\caption{The nominal quantum feedback loop.}
\label{fig:q-robust1}
\end{center}
\end{figure}

Consider the feedback loop shown in Figure \ref{fig:q-robust1}, known as the {\em nominal system}. The network consists
of a quantum component $\Sigma_q$ with gain $g$ (i.e. a mean square inequality of the form \er{gain-q-1} holds). The
algebraic relationship of the network signals is determined by the beamsplitter as \be u_1 = \ep u_0 - \delta y_1 ,
\label{qr1-1} \ee where as usual $\ep^2 + \delta^2=1$. The loop gain is $g \delta$, and so we assume these parameters
have been chosen so that \be g \delta < 1 . \label{qr1-2} \ee By the small gain theorem, this means that this system is
internally stable.

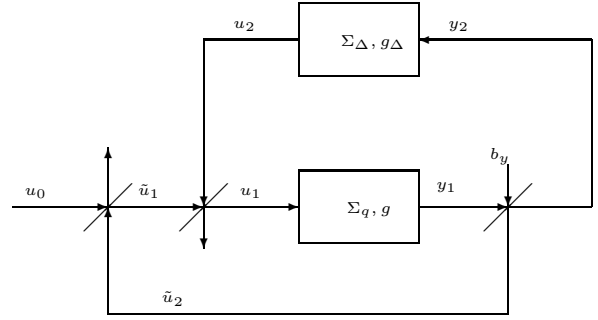
\begin{figure}[h]
\begin{center}
\setlength{\unitlength}{1973sp}%
\begingroup\makeatletter\ifx\SetFigFont\undefined%
\gdef\SetFigFont#1#2#3#4#5{%
  \reset@font\fontsize{#1}{#2pt}%
  \fontfamily{#3}\fontseries{#4}\fontshape{#5}%
  \selectfont}%
\fi\endgroup%
\begin{picture}(7299,3924)(2389,-2473)
\put(8401,-511){\makebox(0,0)[lb]{\smash{{\SetFigFont{6}{7.2}{\familydefault}{\mddefault}{\updefault}{\color[rgb]{0,0,0}$b_y$}%
}}}}
\thinlines
{\color[rgb]{0,0,0}\put(5101,-811){\line(-1,-1){600}}
}%
{\color[rgb]{0,0,0}\put(3601,-1111){\vector( 1, 0){1200}}
}%
{\color[rgb]{0,0,0}\put(4801,-1111){\vector( 1, 0){1200}}
}%
{\color[rgb]{0,0,0}\put(3901,-811){\line(-1,-1){600}}
}%
{\color[rgb]{0,0,0}\put(2401,-1111){\vector( 1, 0){1200}}
}%
{\color[rgb]{0,0,0}\put(8926,-811){\line(-1,-1){600}}
}%
{\color[rgb]{0,0,0}\put(7501,-1111){\vector( 1, 0){1125}}
}%
{\color[rgb]{0,0,0}\put(8626,-1111){\line( 0,-1){1350}}
\put(8626,-2461){\line(-1, 0){5025}}
\put(3601,-2461){\vector( 0, 1){1350}}
}%
{\color[rgb]{0,0,0}\put(6001,539){\framebox(1500,900){}}
}%
{\color[rgb]{0,0,0}\put(8626,-1111){\line( 1, 0){1050}}
\put(9676,-1111){\line( 0, 1){2100}}
\put(9676,989){\vector(-1, 0){2175}}
}%
{\color[rgb]{0,0,0}\put(3601,-1111){\vector( 0, 1){750}}
}%
{\color[rgb]{0,0,0}\put(6001,989){\line(-1, 0){1200}}
\put(4801,989){\vector( 0,-1){2100}}
}%
{\color[rgb]{0,0,0}\put(8626,-586){\vector( 0,-1){525}}
}%
{\color[rgb]{0,0,0}\put(4801,-1111){\vector( 0,-1){525}}
}%
\put(6601,-1186){\makebox(0,0)[lb]{\smash{{\SetFigFont{6}{7.2}{\familydefault}{\mddefault}{\updefault}{\color[rgb]{0,0,0}$\Sigma_q,g$}%
}}}}
\put(7726,-886){\makebox(0,0)[lb]{\smash{{\SetFigFont{6}{7.2}{\familydefault}{\mddefault}{\updefault}{\color[rgb]{0,0,0}$y_1$}%
}}}}
\put(6526,914){\makebox(0,0)[lb]{\smash{{\SetFigFont{6}{7.2}{\familydefault}{\mddefault}{\updefault}{\color[rgb]{0,0,0}$\Sigma_\Delta,g_\Delta$}%
}}}}
\put(5251,-961){\makebox(0,0)[lb]{\smash{{\SetFigFont{6}{7.2}{\familydefault}{\mddefault}{\updefault}{\color[rgb]{0,0,0}$u_1$}%
}}}}
\put(3976,-961){\makebox(0,0)[lb]{\smash{{\SetFigFont{6}{7.2}{\familydefault}{\mddefault}{\updefault}{\color[rgb]{0,0,0}$\tilde u_1$}%
}}}}
\put(5176,1139){\makebox(0,0)[lb]{\smash{{\SetFigFont{6}{7.2}{\familydefault}{\mddefault}{\updefault}{\color[rgb]{0,0,0}$u_2$}%
}}}}
\put(7876,1139){\makebox(0,0)[lb]{\smash{{\SetFigFont{6}{7.2}{\familydefault}{\mddefault}{\updefault}{\color[rgb]{0,0,0}$y_2$}%
}}}}
\put(4276,-2311){\makebox(0,0)[lb]{\smash{{\SetFigFont{6}{7.2}{\familydefault}{\mddefault}{\updefault}{\color[rgb]{0,0,0}$\tilde{u}_2$}%
}}}}
\put(2551,-961){\makebox(0,0)[lb]{\smash{{\SetFigFont{6}{7.2}{\familydefault}{\mddefault}{\updefault}{\color[rgb]{0,0,0}$u_0$}%
}}}}
{\color[rgb]{0,0,0}\put(6001,-1561){\framebox(1500,900){}}
}%
\end{picture}%

\caption{The \lq\lq{actual}\rq\rq \ quantum feedback loop, showing the effect of the environment $\Sigma_\Delta$.}
\label{fig:q-robust1a}
\end{center}
\end{figure}

In reality, such a system will interact with an external environment. One way of modelling the {\em actual system} is
shown in Figure \ref{fig:q-robust1a}. The system $\Sigma_\Delta$ is meant to represent the way in which the environment
interacts with the system $\Sigma_q$, specifically via the feedback loop as shown, in a way
that may depend on the variables of $\Sigma_q$. We assume that $\Sigma_\Delta$ itself is mean square stable with gain
$g_\Delta$. The specific robust stability question here is: does the feedback loop remain stable under the influence of
the environment?

We use the small gain theorem to answer this question. First,
the network signals are related by the equations
\begin{eqnarray}
\tilde{u}_1 &=& \ep u_0 - \delta \tilde{u}_2
\nn \\
u_1 &=& \ep_u \tilde{u}_1 - \delta_u u_2
\nn \\
y_2 &=& \ep_y y_1 - \delta_y b_y
\nn \\
\tilde{u}_2 &=& \delta_y y_1 + \ep_y b_y
\label{qr1-3}
\end{eqnarray}

Next, the gain of the lower part of the network in Figure \ref{fig:q-robust1a}, from input $u_2$ to output $y_2$ can be
determined to be \be g_{u_2 \to y_2}\leq\frac{\delta_u \epsilon_y g}{1-\epsilon_u\delta_y(g\delta)} = g_{max}.
\label{qr1-4} \ee
%In view of \er{qr1-2}, we have $\delta_u \ep_y g < g_{max}$.
The small gain theorem implies the internal stability of the actual quantum feedback loop if \be g_\Delta . g_{u_2 \to
y_2} < 1 , \label{qr1-5} \ee i.e., \be g_\Delta < (g_{max})^{-1} . \label{qr1-5a} \ee Inequality \er{qr1-5a} gives a
bound on the maximum allowable influence of the environment $\Sigma_\Delta$ on the quantum feedback loop that preserves
stability, in terms of $g$, $\delta$, $\ep_u$, $\delta_u$, $\ep_y$, $\delta_y$. We note that if the feedback loop is
removed i.e., let $\delta=0$, so that $\tilde{u}_2$ has no effect on $\Sigma_q$, then \er{qr1-5a} reduces to $g_\Delta
< (\delta_u \ep_y g)^{-1}$, as expected by applying the small gain theorem to the loop formed by $\Sigma_q$ and
$\Sigma_\Delta$.

In general however, the beamsplitter parameters $\ep_u$, $\delta_u$, $\ep_y$, $\delta_y$ specifying how the environment
interacts with the feedback loop may not be known. This can be dealt with by setting a stricter condition than
\er{qr1-5a} on the maximum allowable gain of the environment \be g_\Delta < \frac{1-g\delta}{g} < (g_{max})^{-1},
\label{qr1-6} \ee which depends only on the nominal parameters $g$ and $\delta$. This implies that robust stability is
assured, regardless of beamsplitter parameters, provided the environmental influence has a gain $g_\Delta$ satisfying
condition \er{qr1-6}.

\subsection{Robust Stabilization of an Open Harmonic Oscillator}
\label{sec:osc}

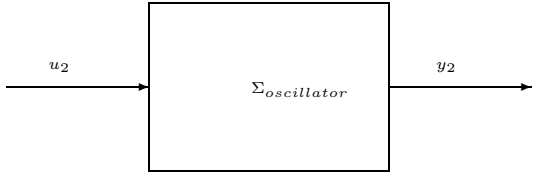
\begin{figure}[h]
\begin{center}
\setlength{\unitlength}{1973sp}%
\begingroup\makeatletter\ifx\SetFigFont\undefined%
\gdef\SetFigFont#1#2#3#4#5{%
  \reset@font\fontsize{#1}{#2pt}%
  \fontfamily{#3}\fontseries{#4}\fontshape{#5}%
  \selectfont}%
\fi\endgroup%
\begin{picture}(6624,2124)(3589,-3973)
\thinlines
{\color[rgb]{0,0,0}\put(5401,-3961){\framebox(3000,2100){}}
}%
{\color[rgb]{0,0,0}\put(3601,-2911){\vector( 1, 0){1800}}
}%
{\color[rgb]{0,0,0}\put(8401,-2911){\vector( 1, 0){1800}}
}%
\put(4126,-2686){\makebox(0,0)[lb]{\smash{{\SetFigFont{6}{7.2}{\familydefault}{\mddefault}{\updefault}{\color[rgb]{0,0,0}$u_2$}%
}}}}
\put(9001,-2686){\makebox(0,0)[lb]{\smash{{\SetFigFont{6}{7.2}{\familydefault}{\mddefault}{\updefault}{\color[rgb]{0,0,0}$y_2$}%
}}}}
\put(6676,-2986){\makebox(0,0)[lb]{\smash{{\SetFigFont{6}{7.2}{\familydefault}{\mddefault}{\updefault}{\color[rgb]{0,0,0}$\Sigma_{oscillator}$}%
}}}}
\end{picture}%

\caption{The open harmonic oscillator, showing the input $u_2$ and output $y_2$ fields which provide the coupling to an
environment $\Sigma_\Delta$ (not shown).} \label{fig:q-osc1}
\end{center}
\end{figure}

In this section we consider the open quantum harmonic oscillator shown in Figure \ref{fig:q-osc1}.

The harmonic oscillator has Hamiltonian \be H= q^2+p^2 , \label{osc-ham} \ee where $p,q$ are the oscillator
quadratures, with the same form as the cavity quadratures defined in Section IIIB2. We have chosen units for which
$\hbar=1$, and set other parameters to unity for simplicity.  The real quadrature of the oscillator, $q$, is coupled to
an external field $u_2$, by an operator $L_2=\sqrt{\kappa}q$. The resulting QLEs for the oscillator quadratures are
\begin{eqnarray}
dq(t) &=& 4p(t) dt
\nn \\
dp(t) &=& -4 q(t) dt  - 2\sqrt{\kappa} dU_{2,i} (t) , \label{osc-1}
\end{eqnarray}
and the commutation relations are, as before, $[q,p]=2i$, $[a,a^\dagger]=1$. These are the equations of motion of a
noisy oscillation, with average motion of frequency $2$ radians per second (when $u_2$ is zero mean). Such models have
been used in the literature e.g., to model an atom trapped in an optical cavity \cite{DJ99,SW06}. It's important to
note that this coupled oscillator has only marginally stable dynamics (its poles lie on the imaginary axis), so it does
not have a finite mean square gain. Therefore this system is very susceptible to environmental influence.

Suppose that the input $u_2$ due to the environment is a displaced field of the form \be dU_2(t)= \beta_{u_2}(t) dt +
dB_{u_2}(t). \label{osc-2} \ee The corresponding output channel $y_2$  is given by \be dY_2(t) = \beta_{y_2}(t) dt +
dB_{u_2}(t), \label{osc-3} \ee where \be \beta_{y_2}(t) = \sqrt{\kappa} q(t) + \beta_{u_2}(t). \label{osc-4} \ee In
addition to the effect of the quantum noise $dB_{u_2}$, we need to address the potentially disruptive influence of the
operator $\beta_{u_2}$, which may depend in some way on oscillator quadratures, via feedback from the environment
$\Sigma_\Delta$. We therefore consider the problem of robustly stabilizing this open oscillator. By this we mean the
problem of constructing a feedback loop that will ensure stability, and tolerate as much environmental influence as
possible.

\begin{figure}[h]
\begin{center}
\setlength{\unitlength}{1973sp}%
\begingroup\makeatletter\ifx\SetFigFont\undefined%
\gdef\SetFigFont#1#2#3#4#5{%
  \reset@font\fontsize{#1}{#2pt}%
  \fontfamily{#3}\fontseries{#4}\fontshape{#5}%
  \selectfont}%
\fi\endgroup%
\begin{picture}(5637,5799)(3676,-4348)
\put(3676,-1636){\makebox(0,0)[lb]{\smash{{\SetFigFont{6}{7.2}{\familydefault}{\mddefault}{\updefault}{\color[rgb]{0,0,0}$u_0$}%
}}}}
\thinlines
{\color[rgb]{0,0,0}\put(6001,989){\line(-1, 0){1200}}
\put(4801,989){\line( 0,-1){1950}}
\put(4801,-961){\vector( 1, 0){1200}}
}%
{\color[rgb]{0,0,0}\put(6001,-1861){\framebox(1500,1200){}}
}%
{\color[rgb]{0,0,0}\put(6001,-3961){\framebox(1500,900){}}
}%
{\color[rgb]{0,0,0}\put(7501,-961){\line( 1, 0){1200}}
\put(8701,-961){\line( 0, 1){1950}}
\put(8701,989){\vector(-1, 0){1200}}
}%
{\color[rgb]{0,0,0}\put(7501,-1561){\line( 1, 0){1200}}
\put(8701,-1561){\line( 0,-1){1950}}
\put(8701,-3511){\vector(-1, 0){1200}}
}%
{\color[rgb]{0,0,0}\put(6001,-3511){\line(-1, 0){1200}}
\put(4801,-3511){\line( 0, 1){1950}}
\put(4801,-1561){\vector( 1, 0){1200}}
}%
{\color[rgb]{0,0,0}\put(4201,-4336){\dashbox{90}(5100,4050){}}
}%
{\color[rgb]{0,0,0}\put(5026,-1336){\line(-1,-1){450}}
}%
{\color[rgb]{0,0,0}\put(4276,-1561){\vector( 1, 0){525}}
}%
{\color[rgb]{0,0,0}\put(4801,-1636){\vector( 0, 1){375}}
}%
\put(7951,-1861){\makebox(0,0)[lb]{\smash{{\SetFigFont{6}{7.2}{\familydefault}{\mddefault}{\updefault}{\color[rgb]{0,0,0}$y_1$}%
}}}}
\put(6226,-3586){\makebox(0,0)[lb]{\smash{{\SetFigFont{6}{7.2}{\rmdefault}{\mddefault}{\updefault}{\color[rgb]{0,0,0}$\Sigma_{controller}$}%
}}}}
\put(6601,914){\makebox(0,0)[lb]{\smash{{\SetFigFont{6}{7.2}{\familydefault}{\mddefault}{\updefault}{\color[rgb]{0,0,0}$\Sigma_\Delta$}%
}}}}
\put(5251,-1936){\makebox(0,0)[lb]{\smash{{\SetFigFont{6}{7.2}{\familydefault}{\mddefault}{\updefault}{\color[rgb]{0,0,0}$u_1$}%
}}}}
\put(6226,-1336){\makebox(0,0)[lb]{\smash{{\SetFigFont{6}{7.2}{\familydefault}{\mddefault}{\updefault}{\color[rgb]{0,0,0}$\Sigma_{oscillator}$}%
}}}}
\put(5251,-1261){\makebox(0,0)[lb]{\smash{{\SetFigFont{6}{7.2}{\familydefault}{\mddefault}{\updefault}{\color[rgb]{0,0,0}$u_2$}%
}}}}
\put(7951,-1261){\makebox(0,0)[lb]{\smash{{\SetFigFont{6}{7.2}{\familydefault}{\mddefault}{\updefault}{\color[rgb]{0,0,0}$y_2$}%
}}}}
{\color[rgb]{0,0,0}\put(6001,539){\framebox(1500,900){}}
}%
\end{picture}%

\caption{The open harmonic oscillator, with coupling to the environment $\Sigma_\Delta$ and quantum feedback control.
The dashed box indicates the feedback loop that we are constructing to improve the robustness of the oscillator.}
\label{fig:q-osc2}
\end{center}
\end{figure}
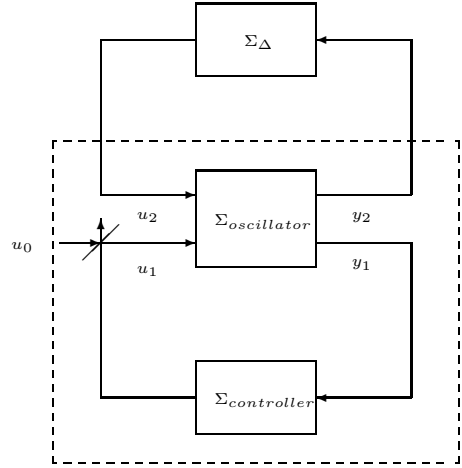

As a first step, we need to establish a channel to mediate the desired feedback, which employs a quantum system as a
controller, and a beamsplitter, as shown in Figure \ref{fig:q-osc2}.

We assume that it is possible to establish a coupling to a second field channel $u_1$ via the operator
$L_1=\sqrt{\gamma} a$, where $2a=q+i p$. The field $u_0$ entering the beamsplitter is assumed to be in the vacuum
state. We seek a controller such that the gain of the oscillator-controller system from $u_2$ to $y_2$ is
\lq\lq{small}\rq\rq--- by the small gain theorem, this allows for a larger  gain of the environment that ensures the
system remains mean square stable. We increase  robustness in this way, by reducing the effect of the environment on
the oscillator.

The equations of motion for the oscillator quadratures coupled to both external fields $u_1,u_2$, but with no
controller present (i.e., no feedback loop, $\beta_{u_1}=0$), are
\begin{eqnarray}
dq(t) &=& (-\frac{\gamma}{2}q(t)+ 4p(t)  )dt - \sqrt{\gamma} dU_{1,r}(t)
\nn \\
dp(t) &=& (-\frac{\gamma}{2}p(t) -4 q(t))dt    - \sqrt{\gamma} dU_{1,i}(t)
\nn \\
&&  \hspace{1.0cm} - 2\sqrt{\kappa} dU_{2,i} (t)
\label{osc-4a}
\end{eqnarray}
It can be seen that the second coupling to $u_1$ provides damping, which stabilizes the oscillator and implies that the
feedback loop has a finite mean square gain. Indeed, we can calculate
\begin{eqnarray}
d \la q^2(t)+p^2(t) \ra &=& \la -\gamma (q^2(t)+p^2(t))
\nn \\
&& - 2 \sqrt{\kappa} (\beta_{u_2,i} (t)p(t)+p(t) \beta_{u_2,i}(t))
\nn \\
&& + \lambda_0 \ra dt
\label{osc-5}
\end{eqnarray}
where $\lambda_0=4 \kappa + 2 \gamma$, which shows the damping effect of this coupling. The gain from $u_2$ to $y_2$
without any controller is estimated from \er{osc-3} and \er{osc-5}  to be \be g_{u_2 \to y_2, \text{(no fb)}} \leq
\frac{\kappa}{\gamma}+1 . \label{osc-6} \ee As expected, this inequality captures the intuitive idea that the effect of
the environment depends on the strength of the two couplings, and in fact is proportional to the ratio $\kappa/\gamma$.
In particular, if the control channel coupling $\gamma$ is small relative to the environment channel coupling $\kappa$,
this gain will be large, which results in low robustness since the maximum gain of the environment allowable for stable
operation will be small.

We now consider the effect of including feedback control.  The output channel $y_1$ corresponding to an input
$dU_1=\beta_{u_1}dt+dB_{u_1}$ is given by \be dY_1(t) = \beta_{y_1}(t)dt + dB_{u_1}(t), \label{osc-7} \ee where \be
\beta_{y_1}(t) = \sqrt{\gamma} a(t) + \beta_{u_1}(t) . \label{osc-8} \ee We choose the controller to be an optical
amplifier or attenuator with gain $g$, and the beamsplitter parameters ($\epsilon,\delta$) such that $\delta g < 1$.
Therefore, using the feedback loop of Figure \ref{fig:q-osc2} and the steady-state model for an amplifier in section
\ref{sec:components-amps} (used here for simplicity) implies that
 \be
 \beta_{u_1} = \delta g \beta_{y_1} .
\label{osc-9} \ee Combining \er{osc-8} and \er{osc-9} we find that \be \beta_{u_1} = \frac{\delta g}{1-\delta
g}\sqrt{\gamma} a = \sqrt{\gamma}G a , \label{osc-10} \ee where $G= \delta g/(1-\delta g)>0$. Thus using convention
\er{quad-beta} \be \beta_{u_1,r} = \sqrt{\gamma} G q, \ \text{and} \  \beta_{u_1,i} = \sqrt{\gamma} G p .
\label{osc-10a} \ee

To see the effect of this feedback, we include the non-zero $\beta_{u_1}$ terms to recalculate (c.f., Eq. \er{osc-5})
\begin{eqnarray}
&& d \la q^2(t)+p^2(t) \ra
\label{osc-11}  \\
 & =& \la -\gamma (q^2(t)+p^2(t))
\nn \\
 &&- 2 \sqrt{\kappa} (\beta_{u_2,i}(t) p(t)+p(t) \beta_{u_2,i}(t))
+ \lambda_2
\nn \\
&&-\sqrt{\gamma} (   \beta_{u_1,r}(t) q(t)+q(t) \beta_{u_1,r}(t)
\nn \\
&&+   \beta_{u_1,i}(t) p(t)+p(t) \beta_{u_1,i}(t)) \ra dt , \nn
\end{eqnarray}
where $\lambda_2$ is a suitable constant. Substituting in the feedback terms from \er{osc-10a} gives
\begin{eqnarray}
d \la q^2(t)+p^2(t) \ra &=& \la -\gamma(1+2G) (q^2(t)+p^2(t))
\nn \\
&& - 2 \sqrt{\kappa} (\beta_{u_2,i} p(t)+p(t) \beta_{u_2,i}(t))
\nn \\
&& + \lambda_2 \ra dt .
\label{osc-12}
\end{eqnarray}
The gain from $u_2$ to $y_2$ with the controller in place is estimated from \er{osc-3} and \er{osc-12}  to be less than
\be g_{u_2 \to y_2, \text{(fb)}} \leq \frac{\kappa}{\gamma(1+2G)}+1 < \frac{\kappa}{\gamma}+1. \label{osc-13} \ee The
feedback increases the effective coupling rate $\gamma$, so that this gain is smaller than the gain without feedback
\er{osc-6}, and therefore improves the robustness of the oscillator against environmental influence.

The design parameters $\delta$ and $g$ can be chosen in some appropriate manner.  Also, a good design will also need to
take into account the effects of added noise, which are reflected in the constant $\lambda_2$. Indeed, one could use a
degenerate parametric amplifier \cite[section 7.2.9]{GZ00} in place of the amplifier used above, to avoid additional
amplifier noise by carefully selecting the appropriate quadrature gains.

The design procedure used here  is a Lyapunov technique which exploits a  {\em passivity} property of the open and
damped oscillator. Such passivity-based control design techniques are well-known in control engineering,  e.g.,
\cite{OSMM01,PF99}, and have been used recently with success in quantum feedback control, e.g., \cite{HSM05}.

We note that it is also possible to consider the use of a classical controller for robust stabilization, although we
don't here.

\section{Conclusions}
\label{sec:conclusion}

In this paper we have demonstrated that the small gain theorem is applicable to the stability analysis of quantum
feedback networks. These networks may include classical components. While we have focused on specific examples
involving quantum optical elements, the general principles should be apparent. We have also applied these principles to
problems of robust stability analysis and design. We expect the small gain theorem and other stability methods will be
useful for the design of quantum technologies. Future work will include further development and application of
stability methods to quantum networks.

\noindent {\bf Acknowledgments.} We wish to thank H.~Wiseman, R.~Van~Handel and I.~Petersen for helpful discussions. This research was supported by the Australian Research Council.

%%%%%%%%%%%%%%%%%%%%%%%%%%%%%%%%%%%%%%%

\bibliographystyle{plain}

%\bibliography{mjbib2004}

\end{document}